\documentclass[11pt]{article}
\usepackage{latexsym}
\usepackage{amsmath,amsfonts,amsbsy,amssymb,amscd}
\usepackage{fancybox,fancyhdr}
\usepackage{cite}
\usepackage{pstricks,pst-node,psfrag}
\usepackage{graphicx,epsfig,picture,graphics}

\psset{unit=1.3cm,linewidth=.5pt,radius=.2}  

\usepackage{float}                          
\usepackage{lscape}                         
\usepackage{bm}
\usepackage{tikz}

\newcommand{\hoch}[1]{$\, ^{#1}$}

\usepackage{color}
\definecolor{MyDarkBlue}{rgb}{0.15,0.15,0.45}    \definecolor{MyGreen}{rgb}{0.15,0.45,0.45}
\definecolor{MyPurple}{rgb}{0.55,0.25,0.55}
\usepackage[linktocpage=true]{hyperref}
\hypersetup{colorlinks=true,citecolor=MyGreen,linkcolor=MyPurple,urlcolor=MyGreen}

\begin{document}

\begin{flushright}
\hfill{ UG-15-59 \\ TUW-15-17}
\end{flushright}
\vskip 2.5cm

\begin{center}
{\Large \bf Newton--Cartan supergravity with torsion \vspace{.4cm}\\ and Schr\"odinger supergravity }
\end{center}
\vspace{27pt}
\begin{center}
{\Large {\bf }}

\vspace{15pt}

{\Large Eric Bergshoeff\hoch{1}, Jan Rosseel\hoch{2}\hoch{3} and Thomas Zojer\hoch{1}}

\vspace{32pt}

\hoch{1} {\it Van Swinderen Institute for Particle Physics and Gravity, University of Groningen,\\
Nijenborgh 4, 9747 AG Groningen, The Netherlands}\\

\vspace{5pt}

\texttt{e.a.bergshoeff@rug.nl, t.zojer@rug.nl} \\

\vspace{10pt}

\hoch{2} {\it Institute for Theoretical Physics, Vienna University of Technology,\\
                               Wiedner Hauptstr.~8--10/136, A-1040 Vienna, Austria}\\

\vspace{5pt}

\texttt{rosseelj@hep.itp.tuwien.ac.at}

\vspace{10pt}

\hoch{3} {\it Albert Einstein Center for Fundamental Physics, University of Bern, \\
Sidlerstrasse 5, 3012 Bern, Switzerland}\\

\vspace{45pt}
\end{center}

\noindent
We derive a torsionfull version of three-dimensional ${\mathcal N}=2$ Newton--Cartan supergravity using a non-relativistic notion of
the superconformal tensor calculus. The ``superconformal'' theory that we start with is Schr\"odinger supergravity which we obtain
by gauging the Schr\"odinger superalgebra. We present two non-relativistic ${\mathcal N}=2$ matter multiplets that can be used as
compensators in the superconformal calculus. They lead to two different off-shell formulations which, in analogy with the
relativistic case, we call ``old minimal'' and ``new minimal'' Newton--Cartan supergravity. We find similarities but also point out
some differences with respect to the relativistic case.

\vspace{15pt}

\thispagestyle{empty}

\vspace{15pt}

 \vfill

\voffset=-40pt

\newpage

\tableofcontents




\section{Introduction \label{sec:intro}}

Recent applications in condensed matter physics and gauge-gravity duality have led to a renewed interest in the question of how
to consistently couple non-relativistic field theories to arbitrary non-relativistic space-time backgrounds.
As in the relativistic case, a consistent coupling of a field theory to arbitrary geometric background data allows one to
covariantly define currents such as the energy-momentum tensor and to study linear response. This geometric approach has been
used in condensed matter physics recently, as a means to construct effective field theories that capture universal properties
of the fractional quantum Hall effect \cite{Hoyos:2011ez,Son:2013rqa,Abanov:2014ula,Gromov:2014vla}. It also plays a prominent
role in recent applications of gauge-gravity duality to condensed matter physics, such as Lifshitz and Schr\"odinger holography
\cite{Kachru:2008yh,Balasubramanian:2008dm,Son:2008ye}. Here, one views non-relativistic conformal field theories as living on
the boundary of a higher-dimensional space-time with non-relativistic isometries, that is a vacuum solution of a suitable dual
gravitational theory. The partition function of the field theory can then be calculated holographically as the partition function
of the dual gravitational theory, in which all fields are subject to well-prescribed fall-off conditions towards the boundary.
The asymptotic values of the fields of the gravitational dual correspond to sources for operators in the conformal field theory
and play the role of arbitrary geometric background data to which the field theory couples.

In both condensed matter and gauge-gravity duality applications, it has been argued that the correct geometric framework to
specify the background data is given by Newton--Cartan geometry with torsion \cite{Gromov:2014vla,Christensen:2013lma,
Christensen:2013rfa,Banerjee:2014nja,Jensen:2014aia,Hartong:2014oma,Hartong:2014pma,Bergshoeff:2014uea,Jensen:2014wha,
Hartong:2015wxa,Geracie:2015dea,Geracie:2015xfa}. Newton--Cartan geometry was first introduced in the context of Newton--Cartan
gravity \cite{Cartan1,Cartan2,Misner:1974qy}, as the differential geometry necessary to cast Newtonian gravity in a covariant
form akin to General Relativity. Even though Newton--Cartan geometry was originally formulated in a metric-like fashion, recent
advances and applications have focused more on an equivalent vielbein formulation, in both torsionless and torsionfull cases. In
this vielbein formulation one introduces temporal and spatial vielbeins that transform under local spatial rotations and Galilean
boosts, as well as spin connections for spatial rotations and Galilean boosts. Crucially, one also includes an extra gauge field
that is associated to particle number conservation. In the torsionless case, the vielbein formulation of Newton--Cartan geometry
can be constructed by gauging the Bargmann algebra, i.e.~the central extension of the Galilei algebra \cite{Andringa:2010it,
Andringa:2013mma}, where the central charge corresponds to particle number. Similarly, it is possible to obtain particular
torsionfull Newton--Cartan geometries by gauging the conformal extension of the Bargmann algebra, namely the Schr\"odinger
algebra \cite{Bergshoeff:2014uea}.

An interesting question is whether Newton--Cartan geometry and
Newton--Cartan gravity can be made compatible with supersymmetry,
i.e.~whether one can construct Newton--Cartan supergravity
theories. Such theories can be relevant for the construction of
supersymmetric non-relativistic field theories, coupled to arbitrary
backgrounds, that could e.g.~be used as toy models to study exact
results in non-relativistic quantum field theory. Relatedly, one might use
Newton--Cartan supergravity theories to see whether localization
techniques, that have proved useful to obtain exact results for
relativistic supersymmetric theories on curved backgrounds
\cite{Pestun:2007rz,Marino:2011nm,Festuccia:2011ws}, can be extended
to non-relativistic theories.

The first example of a Newton--Cartan supergravity theory was obtained
in \cite{Andringa:2013mma} and corresponds to three-dimensional,
$\mathcal{N} = 2$, on-shell, pure Newton--Cartan supergravity with zero
torsion. The independent gauge fields of this theory are given by
\footnote{We use the same notation and conventions as in \cite{Andringa:2013mma}.}
\begin{equation}\label{on-shell}
\textrm{non-relativistic  on-shell}\,:\hskip.5truecm \big(\tau_\mu, e_\mu{}^a, m_\mu, \psi_{\mu \pm}\big)\,.
\end{equation}
Initially, this theory was constructed via a gauging of the $d = 3$,
$\mathcal{N} = 2$ Bargmann superalgebra; it was recently revisited in
\cite{Bergshoeff:2015uaa}, where it was re-obtained from relativistic
$d = 3$, $\mathcal{N} = 2$ supergravity via a procedure that
corresponds to properly taking the non-relativistic limit while keeping an arbitrary frame formulation.
This limiting procedure was then subsequently used to obtain an {\sl off-shell},
pure $d = 3$, $\mathcal{N} = 2$ Newton--Cartan supergravity
theory. Even though these examples show that Newton--Cartan geometry
and gravity can be appropriately supersymmetrized, for practical
purposes it is desirable to construct more elaborate examples than the
pure, torsionless supergravities just mentioned. In particular, in
view of the above mentioned condensed matter and gauge-gravity duality
applications one would like to obtain Newton--Cartan supergravity
theories that include non-trivial torsion as well as matter
couplings. Such theories can generically not be obtained by applying the simple
gauging procedure that led to the on-shell theory of
\cite{Andringa:2013mma}, as not all fields will correspond to gauge
fields of an underlying superalgebra. Since taking a proper and
consistent non-relativistic limit can be rather cumbersome, new
techniques are thus required to obtain such torsionfull and/or
matter-coupled Newton--Cartan supergravity theories.

A very useful way to construct relativistic supergravity theories is
offered by the superconformal tensor calculus (see \cite{Freedman:2012zz}
for an introduction and references). In relativistic superconformal
tensor calculus, one obtains Poincar\'e supergravity theories by
starting from a gauge theory of the superconformal algebra. In
particular, one starts from a so-called `Weyl multiplet', that
realizes the superconformal algebra and contains its gauge fields
(either as independent or as dependent ones). In a next step, one
couples the Weyl multiplet to a `compensator multiplet', whose role is
to gauge fix the superconformal symmetries that are not part of the
Poincar\'e superalgebra. As a concrete example, we remind how
this procedure is applied to obtain $d = 4$, $\mathcal{N} = 1$ `old
minimal' supergravity. In this case, the $d = 4$, $\mathcal{N}=1$ Weyl
multiplet contains the vielbein $E_\mu{}^A$, gravitino $\Psi_\mu$,
$R$-symmetry gauge field $A_\mu$ and dilatation gauge field $b_\mu$ as
independent fields. One can gauge fix the special conformal
transformations by putting $b_\mu$ to zero. As a compensator
multiplet, one takes a chiral multiplet that comprises two complex
scalars $\Phi$ and $F$ and a spinor $\chi$. To derive a Poincar\'e
multiplet from the Weyl multiplet one gauge fixes dilatations,
$R$-symmetry and conformal $S$-supersymmetry. As gauge fixing
conditions, one can choose:
\begin{align}\begin{aligned}\label{relconf}
 \Phi&=1 \,:& \hskip1cm &\textrm{fixes dilatations}\textrm{ and }R\textrm{-symmetry}\,, \\
 \chi&=0 \,:& &\textrm{fixes conformal }S\textrm{-supersymmetry} \,.
\end{aligned}\end{align}
In this way, one obtains the old minimal Poincar\'e multiplet which comprises $(E_\mu{}^A,\Psi_\mu, A_\mu,F)$.
Alternatively, one may also use a tensor multiplet $(\phi, \lambda, B_{\mu\nu})$ as a
compensator multiplet where $\phi$ is a real scalar, $\lambda$ a  spinor and $B_{\mu\nu}$ a 2-form gauge field. Imposing the gauge
fixing conditions
\begin{align}\begin{aligned}\label{relconf2}
 \phi&=1 \,:& \hskip1cm &\textrm{fixes dilatations}\,, \\
 \lambda&=0 \,:& &\textrm{fixes conformal }S\textrm{-supersymmetry} \,,
\end{aligned}\end{align}
one then obtains the new minimal Poincar\'e multiplet with the fields $(E_\mu{}^A,\Psi_\mu,A_\mu, B_{\mu\nu})$.
This theory still enjoys a local $U(1)$-symmetry.

In this paper, we will show that superconformal techniques can also be
used to construct non-relativistic Newton--Cartan supergravity
theories. We will in particular use a non-relativistic analogue of
the superconformal tensor calculus to construct off-shell formulations of
$d=3$, $\mathcal{N}=2$ pure Newton--Cartan supergravity. The
non-relativistic superconformal algebra we will start from is the
Schr\"odinger superalgebra. This algebra contains the Bargmann
superalgebra as a subalgebra (hence our interest in it) and extends it
with a dilatation generator, a single special conformal generator, an
extra bosonic $R$-symmetry generator and a single fermionic $S$-supersymmetry
generator. We will then construct a non-relativistic Schr\"odinger supergravity multiplet
\footnote{We prefer to reserve the name non-relativistic ``conformal'' supergravity multiplet for the multiplet that realizes the
gauging of the Galilean Conformal Superalgebra \cite{deAzcarraga:2009ch,Sakaguchi:2009de,Bagchi:2009ke}. The reason for this is that
the Schr\"odinger superalgebra, with only a single special conformal generator, allows a mass parameter while the Galilean Conformal
Superalgebra does not. We thank Jerzy Lukierski for a discussion on this point.}
that realizes the Schr\"odinger superalgebra and contains its gauge
fields. The independent fields of Schr\"odinger supergravity are  a
temporal vielbein $\tau_\mu$, a spatial vielbein $e_\mu{}^a$, a
central charge gauge field $m_\mu$, a $R$-symmetry gauge field $r_\mu$ and
two gravitini $\psi_{\mu \pm}$. The Schr\"odinger supergravity  multiplet also contains an
extra independent field $b$, that corresponds to the time-like
component of the dilatation gauge field and that can be put to zero by
gauge fixing the special conformal transformation.

In a next step, we will couple the Schr\"odinger supergravity multiplet to a compensator
multiplet, that as in the relativistic case can be used to gauge fix
superfluous superconformal symmetries. We will consider two different
choices of compensator multiplet. The first choice is given by a
non-relativistic $d=3$, $\mathcal{N}=2$ scalar multiplet and this will
lead to a non-relativistic analog of old minimal supergravity with independent fields (see subsection \ref{subsec:old})
\begin{equation}\label{om}
\textrm{non-relativistic old minimal}\,:\hskip.5truecm \big(\tau_\mu,e_\mu{}^a,m_\mu,r_\mu,\psi_{\mu \pm},\chi_-,F_1,F_2\big)\,.
\end{equation}
 The second compensator multiplet we will consider consists of a scalar
$\phi$, a spinor $\lambda$ and an extra bosonic field $S$, that
transforms non-trivially under Galilean boosts. It can be obtained as
a truncation of the non-relativistic limit of a vector multiplet. The
fields $\phi$ and $\lambda$ can then be used to gauge fix dilatations
and $S$-supersymmetry, so that one ends up with a non-relativistic analogue of new minimal
supergravity whose independent fields are given by  (see subsection \ref{subsec:new})
\begin{equation}\label{nm}
\textrm{non-relativistic new minimal}\,:\hskip.5truecm \big(\tau_\mu, e_\mu{}^a, m_\mu, r_\mu, \psi_{\mu \pm}, S\big)\,.
\end{equation}

As was shown in \cite{Bergshoeff:2014uea}, the gauging of the
Schr\"odinger algebra naturally leads to Newton--Cartan geometry with
torsion. The torsion is provided by the spatial components of the
dilatation gauge field, that are dependent on the other fields. This
feature remains in the construction of the Schr\"odinger supergravity multiplet
 and our non-relativistic superconformal
tensor calculus therefore naturally leads to {\sl torsionfull} Newton--Cartan
supergravity theories. In this way, we are thus able to extend the
constructions of \cite{Andringa:2013mma,Bergshoeff:2015uaa} to the
torsionfull case. The torsionless case can be retrieved by putting the
torsion to zero. As the torsion is provided by gauge field components
that depend on the other fields in the supergravity multiplet, this
truncation is non-trivial and its consistency has to be examined. We
will study this truncation in the case of non-relativistic new minimal
supergravity and we will show that this truncation leads to the
off-shell $d=3$, $\mathcal{N}=2$ theory of \cite{Bergshoeff:2015uaa}.

\bigskip\noindent
The organization of this paper is as follows. In section
\ref{sec:schroedinger}, we discuss the gauging of a suitable
Schr\"odinger superalgebra and the ensuing construction of the $d=3$,
$\mathcal{N}=2$ Schr\"odinger supergravity theory. Section \ref{sec:matter} is devoted to
a discussion of the matter multiplets that we will consider as
compensator multiplets. We will show how these multiplets can be
obtained as non-relativistic limits of a relativistic scalar and
vector multiplet and how they can be coupled to the Weyl
multiplet. The construction of torsionfull old minimal and new minimal
Newton--Cartan supergravity will be performed in section \ref{sec:TSNC},
whereas section \ref{sec:notorsion} will be devoted to the truncation
to the torsionless case. Finally, we conclude and give an outlook on
future work in section \ref{sec:conclusions}.


\section{Schr\"odinger supergravity \label{sec:schroedinger}}

In this section we discuss the gauging of superconformal extensions of
the Bargmann algebra, the so-called Schr\"odinger superalgebras. This
is done in several steps. First, in section \ref{subsec:alg} we write
down the transformation rules of all gauge fields, as determined by
the algebra. Then we solve for some of the gauge fields in terms of
others, using so-called conventional curvature constraints. The full
set of curvature constraints is discussed in detail in subsection
\ref{subsec:constr}. Once the dependent gauge fields are expressed in
terms of independent ones, their transformation rules do not
necessarily coincide with those given by the structure constants of
the algebra. The final transformation rules of the dependent gauge
fields thus need to be re-evaluated and this is done in subsection
\ref{subsec:depfields}. Having determined the transformations of all
fields, one can check whether the set of curvature constraints is a
consistent one. This analysis is given in subsection
\ref{subsec:constr} for ease of presentation. Note however that
checking consistency of the constraints constitutes the last step of
the analysis and relies on the transformation rules determined in
subsection \ref{subsec:depfields}.

\subsection{The Schr\"odinger superalgebra and transformation rules \label{subsec:alg}}

Schr\"odinger superalgebras were first found in \cite{Gauntlett:1990xq} as the symmetry group of a spinning particle. However, this
leads to an algebra with a Grassmann valued vector charge, instead of a spinor ($Q_-$ in our notation). Because we are mainly
interested in extensions of the Bargmann superalgebra with two spinorial supercharges we prefer that our Schr\"odinger superalgebra
also contains such operators. For this reason, and because we work in three space-time dimensions, we will work with the
superalgebra of \cite{Leblanc:1992wu}.

For the purpose of this work we restrict ourselves to using $z=2$ Schr\"odinger algebras.
This algebra, as well as its supersymmetric extension, is similar to the Bargmann algebra in that it allows for the same central
extension in the commutator of spatial translations and Galilean boosts. This is important because it enables us to solve for the
non-relativistic spin- and Galilean boost-connections and thus the gauging works in the same way as e.g.~in \cite{Andringa:2010it,
Andringa:2013mma,Bergshoeff:2014uea}.

To be concrete, we use the following set of commutators. The bosonic commutation relations of the Bargmann algebra ($a=1,2$)
\begin{align}\begin{aligned}\label{Bargmannalg}
 \big[P_a,J_{bc}\big] &= 2\,\delta_{a[b}\,P_{c]} \,, &\hskip2cm \big[H,G_a\big] &= P_a \,,  \\
 \big[G_a,J_{bc}\big] &= 2\,\delta_{a[b}\,G_{c]} \,, &  \big[P_a,G_b\big] &= \delta_{ab}\,Z \,,
\end{aligned}\end{align}
are supplemented by the action of the dilatation operator $D$ and special conformal transformations $K$ as follows:
\begin{align}\begin{split}\label{Schrodingeralg}
 \big[D,H\big] &=-2\,H \,, \hskip1cm \big[H,K\big]=D\,, \hskip1.22cm \big[D,K\big] =2\,K \,, \\
 \big[D,P_a\big] &= -P_a \,, \hskip1.06cm \big[D,G_a\big]=G_a \,,\hskip1cm \big[K,P_a\big]=-G_a \,.
\end{split}\end{align}
Here $H,P_a,J_{ab},G_a$ and $Z$ are the generators corresponding to time translations, spatial translations, spatial rotations,
Galilean boosts and central charge transformations, respectively.

The extension to supersymmetry is done by adding two fermionic supersymmetry generators $Q_+$, $Q_-$ and one so-called ``special''
supersymmetry generator $S$. We also have to add one more bosonic so-called $R$-symmetry generator $R$ which, however, does not
contribute to the commutation relations \eqref{Bargmannalg} and \eqref{Schrodingeralg}. This leads to the superalgebra that was
found in \cite{Leblanc:1992wu}, see also \cite{Duval:1993hs,Sakaguchi:2008ku}. In this way the commutators of the Bargmann
superalgebra,
\begin{align}\begin{aligned} \label{3dsuperbargmann}
 \big[J_{ab} , Q_\pm \big] &= -\frac12\,\gamma_{ab}Q_\pm \,, &  \big[G_a , Q_+ \big] &= -\frac12\,\gamma_{a0}Q_- \,,  \\
 \big\{ Q_+ , Q_+ \big\} &= -\gamma^0C^{-1}\,H \,, \hskip1.5cm & \big\{ Q_+ , Q_- \big\} &= -\gamma^aC^{-1}\,P_a \,, \\[.13truecm]
 \big\{ Q_- , Q_- \big\} &= -2\,\gamma^0C^{-1}\,Z \,, & &
\end{aligned}\end{align}
are augmented by the following commutators that involve the extra bosonic and fermionic operators of the Schr\"odinger superalgebra:
\begin{align}\begin{aligned}\label{susySchrodingeralg}
 \big[D,Q_+\big] &=-Q_+ \,, & \big[D,S\big]&=S \,, & \big[R,Q_\pm\big] &= \pm\gamma_0Q_\pm \,, &
   \big[R,S\big] &= \gamma_0S \\
 \big[J_{ab},S\big]&=-\frac12\,\gamma_{ab}S \,,  & \big[S,H\big]&=Q_+ \,, & \big[S,P_a\big] &=\frac12\,\gamma_{a0}Q_-\,, &
   \big[K,Q_+\big] &=S \,, \\
      \hskip1cm \big\{S,S\big\} &= -\gamma^0C^{-1}\,K \,, & & & \big\{S,Q_-\big\} &= \gamma^aC^{-1}\,G_a \,, \end{aligned} \nonumber\\
 \big\{S,Q_+\big\} = \frac12\,\gamma^0C^{-1}\,D+\frac14\,\gamma^{0ab}C^{-1}\,J_{ab} +\frac34\,C^{-1}\,R\,.
              \hspace*{4.95cm}\raisetag{1cm}
\end{align}
According to \cite{Duval:1993hs} this algebra is of a special kind that only exists in odd dimensions. Nevertheless, it will serve
our purpose to  construct a non-relativistic Schr\"odinger supergravity theory  in three dimensions.

After imposing the conventional constraints we will find that the gauge fields $\omega_\mu{}^{ab}$, $\omega_\mu{}^a$, $f_\mu$ and
$\phi_\mu$ of spatial rotations, Galilean boosts, special conformal transformations and $S$-supersymmetry transformations,
respectively, together with the spatial components $b_a=e^\mu{}_ab_\mu$ of the dilatation gauge field $b_\mu$ are dependent. The
time-component $b=\tau^\mu b_\mu$ of $b_\mu$ will turn out to be a St\"uckelberg field for special conformal transformations, just
like in the bosonic case \cite{Bergshoeff:2014uea}. Eventually, we will use this to set $b$ to zero, gauge fixing special conformal
transformations. For notational purposes though, it is easier to keep the full $b_\mu$.

We start with the transformations of the independent bosonic fields under the bosonic symmetries. They are
\begin{align}\begin{split}\label{bostrafo}
 \delta \tau_\mu &= 2\,\Lambda_D\,\tau_\mu \,, \\
 \delta e_\mu{}^a &= \lambda^a{}_b\,e_\mu{}^b +\lambda^a\,\tau_\mu +\Lambda_D\,e_\mu{}^a \,, \\
 \delta m_\mu &= \partial_\mu\sigma +\lambda^a\,e_\mu{}^a \,, \\
 \delta b_\mu &= \partial_\mu\Lambda_D +\Lambda_K\,\tau_\mu \,, \\
 \delta r_\mu &= \partial_\mu \rho \,.
\end{split}\end{align}
For the fermionic fields we find
\begin{align}\begin{split}
 \delta \psi_{\mu+} &= \frac14\,\lambda^{ab}\gamma_{ab}\psi_{\mu+} +\Lambda_D\,\psi_{\mu+} -\gamma_0\psi_{\mu+}\,\rho\,, \\
 \delta \psi_{\mu-} &= \frac14\,\lambda^{ab}\gamma_{ab}\psi_{\mu-} -\frac12\,\lambda^a\gamma_{a0}\psi_{\mu+}
                      +\gamma_0\psi_{\mu-}\,\rho \,.
\end{split}\end{align}
Here $\lambda^a{}_b, \lambda^a, \Lambda_{D}$ and $\rho$ are the parameters of spatial rotations, Galilean boosts, dilatations and
$R$-symmetry transformations, respectively.

The fermionic symmetries act on the bosonic fields as follows:
\begin{align}\begin{split}\label{fermtrafo}
 \delta \tau_\mu &= \frac12\,\bar\epsilon_+\gamma^0\psi_{\mu+} \,, \\
 \delta e_\mu{}^a &= \frac12\,\bar\epsilon_+\gamma^a\psi_{\mu-} +\frac12\,\bar\epsilon_-\gamma^a\psi_{\mu+} \,, \\
 \delta m_\mu &= \bar\epsilon_-\gamma^0\psi_{\mu-} \,, \\
 \delta b_\mu &= -\frac14\,\bar\epsilon_+\gamma^0\phi_\mu -\frac14\,\bar\eta\,\gamma^0\psi_{\mu+} \,, \\
 \delta r_\mu &= -\frac38\,\bar\epsilon_+\phi_\mu +\frac38\,\bar\eta\,\psi_{\mu+} \,,
\end{split}\end{align}
where $\epsilon_{\pm}$ are the two $Q$-supersymmetry parameters while $\eta$ is the single $S$-supersymmetry parameter.
Under these fermionic symmetries  the fermionic fields transform as follows:
\begin{align}\begin{split}
 \delta \psi_{\mu+} &= D_\mu\epsilon_+ -b_\mu\,\epsilon_+ +r_\mu\,\gamma_0\epsilon_+ -\tau_\mu\,\eta \,, \\
 \delta \psi_{\mu-} &= D_\mu\epsilon_-  -r_\mu\,\gamma_0\epsilon_-
                    +\frac12\,\omega_\mu{}^a\gamma_{a0}\epsilon_+     +\frac12\,e_\mu{}^a\gamma_{a0}\eta \,.
\end{split}\end{align}
Since we expect the transformation rules of the dependent gauge fields to change when we solve for them we will not denote them
here. Rather, we will first solve for the gauge fields $\omega_\mu{}^{ab}$, $\omega_\mu{}^a$, $b_a$, $f_\mu$ and $\phi_\mu$,
using conventional curvature constraints. The following subsection is devoted to a discussion of all curvature constraints
of the Schr\"odinger supergravity theory.

\subsection{Curvature constraints \label{subsec:constr}}

While gauging the Schr\"odinger superalgebra we impose several curvature constraints. These follow mostly from requiring the correct
transformation properties under diffeomorphisms. At the same time they allow us to solve for some of the gauge fields in terms of
the remaining independent ones. According to the Schr\"odinger superalgebra the curvatures of the independent gauge fields are given
by
\begin{align}\begin{split}
 \mathcal{R}_{\mu\nu}(H) &= 2\,\partial_{[\mu}\tau_{\nu]} -4\,b_{[\mu}\tau_{\nu]}
          -\frac12\,\bar\psi_{[\mu+}\gamma^0\psi_{\nu]+} \,,\\
 \mathcal{R}_{\mu\nu}{}^a(P) &= 2\,\partial_{[\mu}e_{\nu]}{}^a -2\,\omega_{[\mu}{}^{ab}e_{\nu]}{}^b
          -2\,\omega_{[\mu}{}^a\tau_{\nu]} -2\,b_{[\mu}e_{\nu]}{}^a -\bar\psi_{[\mu+}\gamma^a\psi_{\nu]-} \,, \\[.15truecm]
 \mathcal{R}_{\mu\nu}(Z) &= 2\,\partial_{[\mu}m_{\nu]} -2\,\omega_{[\mu}{}^ae_{\nu]}{}^a -\bar\psi_{[\mu-}\gamma^0\psi_{\nu]-} \,,\\
 \mathcal{R}_{\mu\nu}(D) &= 2\,\partial_{[\mu}b_{\nu]} -2\,f_{[\mu}\tau_{\nu]} +\frac12\,\bar\psi_{[\mu+}\gamma^0\phi_{\nu]} \,, \\
 \mathcal{R}_{\mu\nu}(R) &= 2\,\partial_{[\mu}r_{\nu]} +\frac34\,\bar\psi_{[\mu+}\phi_{\nu]} \,,
\end{split}\end{align}
and
\begin{align}\begin{split}
 \hat\Psi_{\mu\nu+}(Q_+) &= 2\,\partial_{[\mu}\psi_{\nu]+} -\frac12\,\omega_{[\mu}{}^{ab}\gamma_{ab}\psi_{\nu]+}
                     -2\,b_{[\mu}\psi_{\nu]+} +2\,r_{[\mu}\gamma_0\psi_{\nu]+} -2\,\tau_{[\mu}\phi_{\nu]} \,, \\
 \hat\Psi_{\mu\nu-}(Q_-) &= 2\,\partial_{[\mu}\psi_{\nu]-} -\frac12\,\omega_{[\mu}{}^{ab}\gamma_{ab}\psi_{\nu]-}
             -2\,r_{[\mu}\gamma_0\psi_{\nu]-} +\omega_{[\mu}{}^a\gamma_{a0}\psi_{\nu]+} +e_{[\mu}{}^a\gamma_{a0}\phi_{\nu]} \,.
\end{split}\end{align}
The covariant curvatures $\mathcal R$ of the dependent gauge fields are not a priori given by the ``curvatures'' $R$ that follow
from the structure constants of the Schr\"odinger superalgebra since the transformation rules of the dependent gauge fields are not
necessarily equal to the ones that follow from the structure constants of the algebra, see e.g.~the fermionic transformation rules
given in eqs.~\eqref{depfieldfermtrafos}. For the following discussion we will need the curvatures of spatial rotations, Galilean
boosts and $S$-supersymmetry. In the case of spatial rotations the full curvature coincides with the expression that follows from
the structure constants, i.e.~$\mathcal R(J) = R(J)$, but in the other two cases there are additional terms in $\mathcal R$ since
the fermionic transformation rules of those gauge fields contain extra terms beyond those that are determined by the structure
constants, see eq.~\eqref{depfieldfermtrafos}. We therefore have that
\begin{align}\begin{split}\label{depcurv1}
 \mathcal R_{\mu\nu}{}^{ab}(J) &= 2\,\partial_{[\mu}\omega_{\nu]}{}^{ab} -\frac12\,\bar\phi_{[\mu}\gamma^{0ab}\psi_{\nu]+} \,,
\end{split}\end{align}
but that
\begin{align}\label{additional}
\mathcal R_{\mu\nu}{}^a(G) = R_{\mu\nu}{}^a(G) + \textrm{additional terms} \,,
\end{align}
with the structure constant dependent part $R_{\mu\nu}{}^a(G)$ given by
\begin{align}
R_{\mu\nu}{}^a(G) = 2\,\partial_{[\mu}\omega_{\nu]}{}^a -2\,\omega_{[\mu}{}^{ab}\omega_{\nu]}{}^b
             -2\,\omega_{[\mu}{}^a b_{\nu]} -2\,f_{[\mu}e_{\nu]}{}^a +\bar\phi_{[\mu}\gamma^a\psi_{\nu]-} \,.
\end{align}
We will not need the `additional terms' in $\mathcal R(G)$ except for a special trace combination in which case the full expression
for $\mathcal R(G)$ is given by
\begin{align}\label{extra}
\mathcal R_{0a}{}^a(G) = R_{0a}{}^a(G) - e^\mu{}_a\,\bar\psi_{\mu-}\gamma^0\hat\Psi_{a0-}(Q_-)\,.
\end{align}
Using the same notation we find that the curvature of the gauge field of $S$-supersymmetry is given by
\begin{align}
 \mathcal R_{\mu\nu}(S) &= 2\,\partial_{[\mu}\phi_{\nu]} -\frac12\,\omega_{[\mu}{}^{ab}\gamma_{ab}\phi_{\nu]}
                   +2\,b_{[\mu}\,\phi_{\nu]} +2\,r_{[\mu}\,\gamma_0\phi_{\nu]} +2\,f_{[\mu}\,\psi_{\nu]+}   \\
    &\quad +2\,\gamma^0\psi_{[\mu+}\,\Big[\frac14\,\varepsilon^{ab}\,\mathcal R_{\nu]0}{}^{ab}(J) -\mathcal R_{\nu]0}(R)\Big]
           -2\,\gamma^c\psi_{[\mu-}\,\Big[\frac14\,\varepsilon^{ab}\,\mathcal R_{\nu]c}{}^{ab}(J) +\mathcal R_{\nu]c}(R)\Big]
         \nonumber\,,
\end{align}
where the first line comprises all terms that follow from the structure constants.

In the following subsection we will solve for the gauge fields $\omega_\mu{}^{ab}$, $\omega_\mu{}^a$, $b_a$, $f_\mu$ and
$\phi_\mu$ in terms of the independent ones using the following set of conventional constraints:
\begin{align}\begin{aligned}\label{depfieldconstr}
 \mathcal R_{\mu\nu}{}^a(P)&=0 \,,\hskip1cm & \mathcal R_{\mu\nu}(Z) &=0 \,, \hskip1cm& \mathcal R_{a0}(H)&=0 \,, \\
 \hat\Psi_{a0+}(Q_+)&=0 \,, & \gamma^a\hat\Psi_{a0-}(Q_-) &=0 \,, \\
 \mathcal R_{a0}(D) &=0 \,, & \mathcal R_{0a}{}^a(G) &= 0 \,.
\end{aligned}\end{align}
Note that the last constraint involves the curvature of the dependent Galilean boost gauge field whose definition in terms of
the part of the curvature that is determined by the structure constants is given in eq.~\eqref{extra}. Since the conventional
constraints are used to solve for some of the gauge fields their supersymmetry transformations do not lead to new constraints.
We note that, imposing constraints on the curvatures, the Bianchi identities generically imply further constraints on the
curvatures, which holds for the constraints in \eqref{depfieldconstr} and those to be discussed below.

Besides the conventional constraints we also impose the foliation constraint
\begin{equation}\label{RH}
\mathcal R_{\mu\nu}(H)=0\,.
\end{equation}
The time-space component of this constraint is conventional but the space-space part is not. Its $Q_+$-supersymmetry transformation
leads to
\begin{align}\label{Q+=0}
 \hat \Psi_{\mu\nu+}(Q_+) =0 \,,
\end{align}
where, again, only the space-space part is a new, un-conventional constraint. The constraints \eqref{RH} and \eqref{Q+=0} lead to
\begin{align}\label{RD}
 \mathcal R_{ab}(D) =0\,,
\end{align}
as a consequence of a Bianchi identity.
We now consider supersymmetry transformations of the un-conventional constraint $\hat\Psi_{ab+}(Q_+)=0$. A $Q_-$-variation enforces
\footnote{\label{foo} One might wonder how the supersymmetry transformation of a fermionic [bosonic] constraint can lead to
another fermionic [bosonic] constraint. It is true that this is not possible when following generic transformation rules of
covariant quantities. However, those rules only apply if we already know the full set of constraints and the commutator algebra
closes precisely because some constraints are needed to eliminate apparently non-covariant terms. Hence, we can certainly take
guidance from those covariant rules, but when we use them too naively we might miss some constraints.}
\begin{align}\label{Q-ab}
 \hat\Psi_{ab-}(Q_-) =0 \,.
\end{align}
Upon use of all known constraints and Bianchi identities, we find that the only non-trivial variation of \eqref{Q-ab} is its
$Q_-$-variation which we combine with a $Q_+$-variation of \eqref{Q+=0} to find
\begin{align}\label{closureconstr}
 \mathcal R_{ab}(R)=0 \,, \hskip1.5cm \mathcal R_{ab}{}^{cd}(J) =0 \,.
\end{align}
At this point we have checked the symmetry variations of all constraints except the last two, i.e.~\eqref{closureconstr}. Before we
go on determining the implications of their transformations we note that using all constraints so far we find the Bianchi identity
\begin{align}\label{RS=0}
 \mathcal R_{ab}(S) =0 \,.
\end{align}
The only non-trivial transformation of $\mathcal R_{ab}(R)=0$ then leads to \textsuperscript{\ref{foo}}
\begin{align}\label{RJRR}
 \frac34\,\varepsilon^{ab}\,\mathcal R_{\mu\nu}{}^{ab}(J) = \mathcal R_{\mu\nu}(R) \,.
\end{align}
Since \eqref{RJRR} essentially identifies $\mathcal R(J)$ with $\mathcal R(R)$ we have derived all consequences of
\eqref{closureconstr}. The constraint \eqref{RJRR} itself is inert under all symmetries and hence we have derived the full set of
un-conventional constraints that follow from \eqref{RH}.

In summary, the set of constraints comprises the following chain of un-conventional constraints:
\begin{align}\begin{split}\label{unconvconstr}
 \mathcal R_{ab}(H)=0 \quad\stackrel{Q_+}{\longrightarrow}\quad \hat\Psi_{ab+}&=0
    \quad\stackrel{Q_-}{\longrightarrow}\quad \hat\Psi_{ab-}=0  \quad\longrightarrow\\
 \begin{matrix} \hat\Psi_{ab+}=0 \\ \hat\Psi_{ab-}=0 \end{matrix} \,\bigg\}
     \quad\stackrel{Q_\pm}{\longrightarrow}\quad \mathcal R_{ab}(R)&=0  \quad\stackrel{Q_+}{\longrightarrow}\quad
                   \frac34\,\varepsilon^{ab}\,\mathcal R_{\mu\nu}{}^{ab}(J) = \mathcal R_{\mu\nu}(R) \,.
\end{split}\end{align}
The Bianchi identities that feature in the discussion above are given by
\begin{align}\label{Bianchis}
 \mathcal R_{ab}(D) =0 \,, \hskip.8cm \mathcal R_{0[a}{}^{b]}(G) =0 \,, \hskip.8cm
   \mathcal R_{ab}{}^c(G) = 2\,\mathcal R_{0[a}{}^{b]c}(J) \,, \hskip.8cm \mathcal R_{ab}(S)=0 \,.
\end{align}

\subsection{The dependent gauge fields \label{subsec:depfields}}

Let us now determine the expressions of the dependent gauge fields. We first determine the spatial component of $b_\mu$. Using
$\mathcal{R}_{a0}(H)=0$ we find
\begin{align}\label{ba}
 b_a=e^\mu{}_ab_\mu = \frac12\,e^\mu{}_a\tau^\nu\,\big(2\,\partial_{[\mu}\tau_{\nu]}
                                             -\frac12\,\bar\psi_{[\mu+}\gamma^0\psi_{\nu]+}\big) \,.
\end{align}
The (independent) scalar $b=\tau^\mu b_\mu$ is a St\"uckelberg field for special conformal transformations:
\begin{align}\begin{split}
 \delta b &= \Lambda_K +\tau^\mu\partial_\mu\Lambda_D -2\,\Lambda_D\,b -\lambda^a\,b_a
            -\frac14\,\tau^\mu\,(\bar\epsilon_+\gamma^0\phi_\mu+\bar\eta\,\gamma^0\psi_{\mu+}) \\
     &\quad -\frac12\,b\,\bar\epsilon_+\gamma^0\psi_{\rho+}\tau^\rho
            -\frac12\,b_a\,\tau^\rho\,(\bar\epsilon_+\gamma^a\psi_{\rho-}+\bar\epsilon_-\gamma^a\psi_{\rho+}) \,.
\end{split}\end{align}
Thus, we could choose to set $b=0$. This would induce the compensating transformation
\begin{align}\begin{split}\label{lambdaKcomp}
 \Lambda_K &= -\tau^\mu\partial_\mu\Lambda_D +\lambda^a\,b_a
           +\frac14\,\tau^\mu\,(\bar\epsilon_+\gamma^0\phi_\mu+\bar\eta\,\gamma^0\psi_{\mu+}) \\
   &\quad  +\frac12\,b_a\,\tau^\rho\,(\bar\epsilon_+\gamma^a\psi_{\rho-}+\bar\epsilon_-\gamma^a\psi_{\rho+}) \,.
\end{split}\end{align}
In the following we will keep $b\ne 0$. In any case, since no independent field transforms under special conformal transformations
there is in essence no effect from this gauge fixing.

We proceed with determining the other dependent gauge fields. The gauge fields  $\omega_\mu{}^{ab}$ of spatial rotations and
$\omega_\mu{}^a$ of Galilean boosts are solved for using the conventional constraints $\mathcal R_{\mu\nu}{}^a(P)=0$ and
$\mathcal R_{\mu\nu}(Z) =0$. We find the following expressions:
\begin{align}
 \omega_\mu{}^{ab} &= 2\,e^{\nu[a}\big(\partial_{[\nu}e_{\mu]}{}^{b]} -\frac12\,\psi_{[\nu+}\gamma^{b]}\psi_{\mu]-}
                -b_{[\nu}\,e_{\mu]}{}^{b]}\big) \\
    &\quad +e_\mu{}^ce^{\rho a}e^{\nu b}\big(\partial_{[\rho}e_{\nu]}{}^c -\frac12\,\psi_{[\rho+}\gamma^c\psi_{\nu]-}
                -b_{[\rho}\,e_{\nu]}{}^c\big)
   -\tau_\mu e^{\rho a}e^{\nu b}\big(\partial_{[\rho}m_{\nu]}-\frac12\,\psi_{[\rho-}\gamma^0\psi_{\nu]-}\big) \,,
\nonumber\\
\begin{split}
 \omega_\mu{}^a &= -\tau^\nu \big(\partial_{[\nu}e_{\mu]}{}^a -\frac12\,\psi_{[\nu+}\gamma^a\psi_{\mu]-}
                -b_{[\nu}\,e_{\mu]}{}^a\big)  \\
    &\quad +e_\mu{}^ce^{\rho a}\tau^\nu \big(\partial_{[\rho}e_{\nu]}{}^c -\frac12\,\psi_{[\rho+}\gamma^c\psi_{\nu]-}
                -b_{[\rho}\,e_{\nu]}{}^c\big) \\
    &\quad +e^{\nu a}\big(\partial_{[\mu}m_{\nu]}-\frac12\,\psi_{[\mu-}\gamma^0\psi_{\nu]-}\big)
          -\tau_\mu e^{\rho a}\tau^\nu\big(\partial_{[\rho}m_{\nu]}-\frac12\,\psi_{[\rho-}\gamma^0\psi_{\nu]-}\big) \,.
\end{split}
\end{align}
The $S$-supersymmetry gauge field $\phi_\mu$ is determined through the conventional constraints $\hat\Psi_{a0+}(Q_+)=0$
and $\gamma^a\hat\Psi_{a0-}(Q_-) =0$, which lead
to the following expression:
\begin{align}\begin{split}
 \phi_\mu &= -\tau^\nu \,\big(2\,\partial_{[\mu}\psi_{\nu]+}
                 -\frac12\,\omega_{[\mu}{}^{ab}\gamma_{ab}\psi_{\nu]+} -2\,b_{[\mu}\psi_{\nu]+}
                 +2\,r_{[\mu}\gamma_0\psi_{\nu]+} \big) \\
     &\quad  +\tau_\mu\tau^\rho e^\nu{}_c\gamma^{0c}\,\big(2\,\partial_{[\rho}\psi_{\nu]-}
                 -\frac12\,\omega_{[\rho}{}^{ab}\gamma_{ab}\psi_{\nu]-} +\omega_{[\rho}{}^a\gamma_{a0}\psi_{\nu]+}
                 -2\,r_{[\mu}\gamma_0\psi_{\nu]-}\big) \,.
\end{split}\end{align}
Finally, to solve for the special conformal boost gauge field $f_\mu$ we use the conventional constraints $\mathcal R_{a0}(D)=0$
and $\mathcal R_{0a}{}^a(G) =0$. In this way we find that
\begin{align}\label{fmu}
 f_\mu &= \tau^\nu\,\big(2\,\partial_{[\mu}b_{\nu]} +\frac12\,\bar\psi_{[\mu+}\gamma^0\phi_{\nu]}\big) \\
    &\quad     +\frac12\,\tau_\mu\tau^\rho e^\nu{}_a \,\big(2\,\partial_{[\rho}\omega_{\nu]}{}^a
       -2\,\omega_{[\rho}{}^{ab}\omega_{\nu]}{}^b -2\,\omega_{[\rho}{}^ab_{\nu]} +\bar\phi_{[\rho}\gamma^a\psi_{\nu]-}\big)
       -\frac12\,\tau_\mu e^\rho{}_a\,\bar\psi_{\rho-}\gamma^0\hat\Psi_{a0-}(Q_-)\,.   \nonumber
\end{align}

At this point we have solved for all the dependent gauge fields in terms of the independent ones. Using their expressions in terms
of the independent gauge fields, we find that they transform under the bosonic Schr\"odinger transformations as follows:
\begin{align}\begin{split}\label{depfieldbostrafos}
 \delta \omega_\mu{}^{ab} &= \partial_\mu\lambda^{ab} \,, \\
 \delta \omega_\mu{}^a &= \partial_\mu\lambda^a -\omega_\mu{}^a{}_b\lambda^b +b_\mu\lambda^a+\lambda^a{}_b\,\omega_\mu{}^b
               -\Lambda_D\,\omega_\mu{}^a +\Lambda_K\,e_\mu{}^a \,, \\
 \delta f_\mu &= \partial_\mu\Lambda_K +2\,\Lambda_K\,b_\mu -2\,\Lambda_D \,f_\mu
               -\tau_\mu\,\lambda^b\,\mathcal R_{0a}{}^{ab}(J)\,, \\
 \delta \phi_\mu &= \frac14\,\lambda^{ab}\gamma_{ab}\phi_\mu -\Lambda_D\,\phi_\mu -\Lambda_K\,\psi_{\mu+}
               -\gamma_0\phi_\mu\,\rho \,.
\end{split}\end{align}
These are precisely the transformation rules that follow from the structure constants of the Schr\"odinger algebra except for the
curvature term in the transformation rule of the special conformal boost gauge field $f_\mu$. In \cite{Bergshoeff:2014uea} this was
circumvented by redefining $f_\mu$ by adding terms with $m_\mu$ and $\mathcal R_{\mu\nu}{}^{ab}(J)$ in the conventional constraint
$\mathcal R_{0a}{}^a(G)=0$ that is used to solve for $f_\mu$. However, then the field acquired a non-trivial transformation under
the central charge symmetry. We will not perform any redefinition of that kind here.

Concerning the fermionic symmetries, we find that the $Q$ and $S$-transformations of the  dependent gauge fields fields
$\omega_\mu{}^{ab}, \omega_\mu{}^a$ and $\phi_\mu$ are given by
\begin{align}\begin{split}\label{depfieldfermtrafos}
 \delta\omega_\mu{}^{ab} &=-\frac14\,\bar\epsilon_+\gamma^{ab0}\phi_\mu   +\frac14\,\bar\eta\,\gamma^{ab0}\psi_{\mu+} \,, \\
 \delta\omega_\mu{}^a &= \bar\epsilon_-\gamma^0\hat\Psi_\mu{}^a{}_-(Q_-)  -\frac12\,\bar\epsilon_-\gamma^a\phi_\mu
         +\frac14\,e_{\mu b}\,\bar\epsilon_+\gamma^b\hat\Psi^a{}_{0-}(Q_-)  \\
  &\quad +\frac14\,\bar\epsilon_+\gamma^a\hat\Psi_{\mu0-}(Q_-)   -\frac12\,\bar\eta\,\gamma^a\psi_{\mu-}               \,, \\
 \delta\phi_\mu &= D_\mu\eta +b_\mu\,\eta +r_\mu\,\gamma_0\eta +f_\mu\,\epsilon_+ \\
  &\quad +\gamma_0\epsilon_+\,\Big[\frac14\,\varepsilon^{ab}\,\mathcal R_{\mu0}{}^{ab}(J) -\mathcal R_{\mu0}(R)\Big]
         +\gamma^c\epsilon_-\,\Big[\frac14\,\varepsilon^{ab}\,\mathcal R_{\mu c}{}^{ab}(J) +\mathcal R_{\mu c}(R)\Big] \,.
\end{split}\end{align}

The above bosonic and fermionic  transformations allow us to explicitly check that the commutator algebra of two supersymmetries is
realized by the formula
\begin{align}\begin{split}
 \big[\delta(Q_1,S_1),\delta(Q_2,S_2)\big] &= \delta_{\rm g.c.t.}\big(\Xi^\mu\big) +\delta_J\big(\Lambda^{ab}\big)
         +\delta_G\big(\Lambda^a\big) +\delta_Z\big(\Sigma\big) +\delta_D\big(\lambda_D\big) \\
  &\quad +\delta_K\big(\lambda_K\big) +\delta_{Q_+}\big(\Upsilon_+\big) +\delta_{Q_-}\big(\Upsilon_-\big)
         +\delta_S\big(\eta\big) +\delta_R\big(\rho_R\big) \,,
\end{split}\end{align}
where the parameters are given by
\begin{align}\begin{aligned}\label{softalgparam}
 \Xi^\mu &= \frac12\,\bar\epsilon_{2+}\gamma^0\epsilon_{1+}\,\tau^\mu
   +\frac12\big(\bar\epsilon_{2+}\gamma^a\epsilon_{1-}+\bar\epsilon_{2-}\gamma^a\epsilon_{1+}\big)e^\mu{}_a \,, \\
 \Lambda^{ab} &= -\Xi^\mu\omega_\mu{}^{ab} +\frac14\big(\bar\epsilon_{1+}\gamma^{0ab}\eta_2
                                                    -\bar\eta_1\,\gamma^{0ab}\epsilon_{2+}\big)\,, &
        \Upsilon_\pm &= -\Xi^\mu\psi_{\mu\pm} \,, \\
 \Lambda^a &= -\Xi^\mu\omega_\mu{}^a -\frac12\big(\bar\epsilon_{1-}\gamma^a\eta_2 +\bar\eta_1\,\gamma^a\epsilon_{2-}\big)\,, &
        \lambda_K &= -\Xi^\mu f_\mu +\frac12\,\bar\eta_2\,\gamma^0\eta_1 \,, \\
 \Sigma &= -\Xi^\mu m_\mu +\bar\epsilon_{2-}\gamma^0\epsilon_{1-} \,, &
        \rho_R &= -\Xi^\mu r_\mu +\frac38\big(\bar\epsilon_{1+}\eta_2 -\bar\eta_1\,\epsilon_{2+}\big)\,, \\
 \lambda_D &= -\Xi^\mu b_\mu +\frac14\big(\bar\epsilon_{1+}\gamma^0\eta_2 +\bar\eta_1\,\gamma^0\epsilon_{2+}\big)\,, &
        \eta &= -\Xi^\mu\phi_\mu \,.       \raisetag{3cm}
\end{aligned}\end{align}
This finishes the discussion of the Schr\"odinger supergravity theory.

Note that our analysis of the Schr\"odinger theory is not fully complete, since we did not derive the variation of the dependent
field $f_\mu$ under fermionic symmetries. Even so, this was not needed to show that the set of constraints \eqref{unconvconstr}
is a consistent one and that the commutator algebra closes on all independent fields.


\section{Matter multiplets \label{sec:matter}}

In this section we present matter multiplets that realize the same commutators corresponding to the Schr\"odinger superalgebra as
we derived for the Schr\"odinger supergravity multiplet in the previous section. These multiplets will be used as compensator
multiplets in the next section to derive off-shell formulations of Newton--Cartan supergravity.

One such off-shell formulation already exists in the literature \cite{Bergshoeff:2015uaa}. It was obtained by taking a
non-relativistic limit of the three-dimensional $\mathcal{N}=2$ new minimal Poincar\'e multiplet \cite{Howe:1995zm}. The new minimal
Poincar\'e multiplet follows from superconformal techniques using a compensating (relativistic) vector multiplet. Hence, in order to
derive its non-relativistic analog we should use as a compensator a non-relativistic vector multiplet. This is one of the two
non-relativistic matter multiplets which we derive in this section. The other one is the scalar multiplet which we shall later use
to derive a new off-shell formulation of Newton--Cartan supergravity.

It would be very efficient if we could derive the matter multiplets coupled to Schr\"odinger supergravity by applying the
non-relativistic limiting procedure of \cite{Bergshoeff:2015uaa}. However we cannot, because the Schr\"odinger superalgebra does not
follow from the contraction of any relativistic superalgebra and the same applies to the corresponding Schr\"odinger supergravity
theory. Instead, we shall start from the rigid version of a relativistic matter multiplet that realizes the Poincar\'e superalgebra.
First, we use that as a starting point to derive a non-relativistic matter multiplet that realizes the rigid Bargmann superalgebra.
\footnote{This (rigid) limit  coincides with the non-relativistic limit performed in \cite{Gomis:2004pw}.}
The important thing is that we have now derived the field content of the non-relativistic multiplet. It turns out that the same
multiplet also provides a representation of the rigid Schr\"odinger superalgebra. Therefore, once we have obtained this
non-relativistic multiplet, we can couple it to the fields of Schr\"odinger supergravity, thereby realizing the commutator algebra
derived in the previous section, in the standard way.

\subsection{The scalar multiplet \label{subsec:matter}}

In this subsection we construct the non-relativistic scalar multiplet. We start with the three-dimensional rigid relativistic
$\mathcal{N}=2$ scalar multiplet which comprises two complex scalars and two spinors. In real notation we are thus left with the
fields $(\varphi_1,\varphi_2,\chi_1,\chi_2,F_1,F_2)$:
\begin{align}\begin{split}\label{relscalar}
 \delta \varphi_1 &= \bar\eta_1\chi_1 +\bar\eta_2\chi_2 \,, \\[.15truecm]
 \delta \varphi_2 &= \bar\eta_1\chi_2 -\bar\eta_2\chi_1 \,, \\
 \delta \chi_1 &= \frac14\,\gamma^\mu \partial_\mu \varphi_1\,\eta_1   -\frac14\,\gamma^\mu \partial_\mu \varphi_2\,\eta_2
                 -\frac14\,F_1\,\eta_1   -\frac14\,F_2\,\eta_2 \,, \\
 \delta \chi_2 &= \frac14\,\gamma^\mu \partial_\mu \varphi_2\,\eta_1   +\frac14\,\gamma^\mu \partial_\mu \varphi_1\,\eta_2
                 -\frac14\,F_2\,\eta_1   +\frac14\,F_1\,\eta_2 \,, \\[.1truecm]
 \delta F_1 &= -\bar\eta_1\gamma^\mu \partial_\mu \chi_1 +\bar\eta_2\gamma^\mu \partial_\mu\chi_2 \,, \\[.15truecm]
 \delta F_2 &= -\eta_1\gamma^\mu \partial_\mu\chi_2 -\eta_2\gamma^\mu \partial_\mu \chi_1 \,.
\end{split}\end{align}
To take the non-relativistic limit we use a contraction parameter $\omega$ which we will send to infinity. The rescaling of the
symmetry parameters follows from the In\"on\"u--Wigner contraction of the related symmetry generators, see \cite{Bergshoeff:2015uaa}.
This means for example that we will require
\begin{align}\label{espscale}
 \epsilon_\pm = \frac{\omega^{\mp1/2}}{\sqrt 2}\,\big(\eta_1 \pm\gamma_0\eta_2\big) \,.
\end{align}
It remains to find the scalings of all other fields. It turns out that, in order to avoid terms that diverge in the limit
$\omega\to\infty$, we need to use
\begin{align}\label{chiscale}
 \chi_\pm = \frac{\omega^{-1\pm1/2}}{\sqrt 2}\,(\chi_1 \pm\gamma_0\chi_2)
\end{align}
for the two spinors, while for the scalings of the bosons we need to take
\begin{align}\label{phifscale}
 \tilde \varphi_i = \frac1\omega\,\varphi_i \,, \hskip2cm \tilde F_i = -\frac1\omega \,F_i \,.
\end{align}
After calculating the transformation rules in the limit $\omega\to\infty$ we drop the tildes and find
\begin{align}\begin{split}\label{rigidscalar}
 \delta \varphi_1 &= \bar\epsilon_+\chi_+  +\bar\epsilon_-\chi_- \,, \\[.15truecm]
 \delta \varphi_2 &= \bar\epsilon_+\gamma^0\chi_+   -\bar\epsilon_-\gamma^0\chi_- \,, \\
 \delta \chi_+ &= \frac14\,\gamma^0\epsilon_+\,\partial_t\varphi_1   +\frac14\,\epsilon_+\,\partial_t\varphi_2
             +\frac14\,\gamma^i\epsilon_-\,\partial_i\varphi_1  +\frac14\,\gamma^{i0}\epsilon_-\,\partial_i\varphi_2
             +\frac14\,\epsilon_-\,F_1    +\frac14\,\gamma_0\epsilon_-\,F_2 \,, \\
 \delta \chi_- &= \frac14\,\gamma^i\epsilon_+\,\partial_i\varphi_1  -\frac14\,\gamma^{i0}\epsilon_+\,\partial_i\varphi_2
             +\frac14\,\epsilon_+\,F_1 -\frac14\,\gamma_0\epsilon_+\,F_2 \,, \\[.15cm]
 \delta F_1 &= \bar\epsilon_+\gamma^i\partial_i\chi_+  +\bar\epsilon_+\gamma^0\partial_t\chi_-
             +\bar\epsilon_-\gamma^i\partial_i\chi_-  \,, \\[.2cm]
 \delta F_2 &= \bar\epsilon_+\gamma^{i0}\partial_i\chi_+  +\bar\epsilon_+\partial_t\chi_-
             -\bar\epsilon_-\gamma^{i0}\partial_i\chi_- \,.
\end{split}\end{align}
Together with the bosonic transformation rules, which we refrain from giving here but which can be obtained easily by similar
techniques, the transformation rules \eqref{rigidscalar} realize the rigid Bargmann superalgebra. Next, we promote this multiplet
to a representation of the rigid Schr\"odinger superalgebra by assigning transformations under the Schr\"odinger transformations
that are not contained in the Bargmann superalgebra. After that we couple the multiplet to the fields of Schr\"odinger supergravity.
Following standard techniques of coupling matter to supergravity we find for the bosonic transformations
\begin{align}\begin{split}\label{schroscalarbos}
 \delta \varphi_1 &= w\,\Lambda_D\varphi_1  +\frac{2w}{3}\,\rho\,\varphi_2 \,, \\
 \delta \varphi_2 &= w\,\Lambda_D\varphi_2  -\frac{2w}{3}\,\rho\,\varphi_1 \,, \\
 \delta \chi_+ &= \frac14\,\lambda^{ab}\gamma_{ab}\chi_+   -\frac12\,\lambda^a\gamma_{a0}\chi_-  +(w-1)\,\Lambda_D\chi_+
             -\big(\frac{2w}{3}+1\big)\,\gamma_0\chi_+\,\rho \,, \\
 \delta \chi_- &= \frac14\,\lambda^{ab}\gamma_{ab}\chi_-  +w\,\Lambda_D\chi_-
             +\big(\frac{2w}{3}+1\big)\,\gamma_0\chi_-\,\rho   \,,  \\[.15cm]
 \delta F_1 &= (w-1)\,\Lambda_DF_1 +2\,\big(\frac{w}{3}+1\big)\,\rho\,F_2 \,, \\[.2cm]
 \delta F_2 &= (w-1)\,\Lambda_DF_2 -2\,\big(\frac{w}{3}+1\big)\,\rho\,F_1\,,
\end{split}\end{align}
while for the fermionic transformation rules we find the following expressions:
\begin{align}\begin{split}\label{schroscalarferm}
 \delta \varphi_1 &= \bar\epsilon_+\chi_+  +\bar\epsilon_-\chi_- \,, \\[.2cm]
 \delta \varphi_2 &= \bar\epsilon_+\gamma^0\chi_+   -\bar\epsilon_-\gamma^0\chi_- \,, \\[.15cm]
 \delta \chi_+ &= \frac14\,\gamma^0\epsilon_+\,\tau^\mu\hat D_\mu\varphi_1   +\frac14\,\epsilon_+\,\tau^\mu\hat D_\mu\varphi_2
             +\frac14\,\gamma^a\epsilon_-\,e^\mu{}_a\hat D_\mu\varphi_1
             +\frac14\,\gamma^{a0}\epsilon_-\,e^\mu{}_a\hat D_\mu\varphi_2 \\
      &\quad +\frac14\,\epsilon_-\,F_1    +\frac14\,\gamma_0\epsilon_-\,F_2
             -\frac{w}{4}\,\gamma^0\eta\,\varphi_1 -\frac{w}{4}\,\eta\,\varphi_2 \,, \\
 \delta \chi_- &= \frac14\,\gamma^a\epsilon_+\,e^\mu{}_a\hat D_\mu\varphi_1
             -\frac14\,\gamma^{a0}\epsilon_+\,e^\mu{}_a\hat D_\mu\varphi_2
             +\frac14\,\epsilon_+\,F_1 -\frac14\,\gamma_0\epsilon_+\,F_2 \,, \\[.15cm]
 \delta F_1 &= \bar\epsilon_+\gamma^ae^\mu{}_a\hat D_\mu\chi_+  +\bar\epsilon_+\gamma^0\tau^\mu\hat D_\mu\chi_-
             +\bar\epsilon_-\gamma^ae^\mu{}_a\hat D_\mu\chi_- -(w+1)\,\bar\eta\,\gamma^0\chi_- \,, \\[.2cm]
 \delta F_2 &= \bar\epsilon_+\gamma^{a0}e^\mu{}_a\hat D_\mu\chi_+  +\bar\epsilon_+\tau^\mu\hat D_\mu\chi_-
             -\bar\epsilon_-\gamma^{a0}e^\mu{}_a\hat D_\mu\chi_- -(w+1)\,\bar\eta\,\chi_-  \,.
\end{split}\end{align}
The covariant derivatives that appear in \eqref{schroscalarferm} can be deduced from the transformation rules \eqref{schroscalarbos}
and \eqref{schroscalarferm}. For the bosonic fields they are given by
\begin{align}\begin{split}
 \hat D_\mu\varphi_1 &= \partial_\mu\varphi_1 -w\,b_\mu\,\varphi_1 -\frac{2w}{3}\,r_\mu\,\varphi_2
          -\bar\psi_{\mu+}\chi_+  -\bar\psi_{\mu-}\chi_- \,, \\
 \hat D_\mu\varphi_2 &= \partial_\mu\varphi_2 -w\,b_\mu\,\varphi_2 +\frac{2w}{3}\,r_\mu\,\varphi_1
          -\bar\psi_{\mu+}\gamma^0\chi_+  +\bar\psi_{\mu-}\gamma^0\chi_- \,, \\
 \hat D_\mu F_1 &= \partial_\mu F_1 -(w-1)\,b_\mu\,F_1 -2\,\big(\frac{w}{3}+1\big)\,r_\mu\,F_2 \\
     &\quad -\bar\psi_{\mu+}\gamma^ae^\rho{}_a\hat D_\rho\chi_+  -\bar\psi_{\mu+}\gamma^0\tau^\rho\hat D_\rho\chi_-
          -\bar\psi_{\mu-}\gamma^ae^\rho{}_a\hat D_\rho\chi_-  +(w+1)\,\bar\phi_\mu\gamma^0\chi_- \,, \\
 \hat D_\mu F_2 &= \partial_\mu F_2 -(w-1)\,b_\mu\,F_2 +2\,\big(\frac{w}{3}+1\big)\,r_\mu\,F_1 \\
     &\quad -\bar\psi_{\mu+}\gamma^{a0}e^\rho{}_a\hat D_\rho\chi_+ -\bar\psi_{\mu+}\tau^\rho\hat D_\rho\chi_-
          +\bar\psi_{\mu-}\gamma^{a0}e^\rho{}_a\hat D_\rho\chi_-  +(w+1)\,\bar\phi_\mu\chi_- \,,
\end{split}\end{align}
while for the covariant derivatives of the fermions we find the following expressions:
\begin{align}\begin{split}
 \hat D_\mu\chi_+ &= D_\mu\chi_+  +\frac12\,\omega_\mu{}^a\gamma_{a0}\chi_- -(w-1)\,b_\mu\chi_+
          +\big(\frac{2w}{3}+1\big)\,r_\mu\,\gamma_0\chi_+ \\
     &\quad -\frac14\,\gamma^0\psi_{\mu+}\,\tau^\rho\hat D_\rho\varphi_1
          -\frac14\,\psi_{\mu+}\,\tau^\rho\hat D_\rho\varphi_2  -\frac14\,\gamma^a\psi_{\mu-}\,e^\rho{}_a\hat D_\rho\varphi_1 \\
     &\quad -\frac14\,\gamma^{a0}\psi_{\mu-}\,e^\rho{}_a\hat D_\rho\varphi_2  -\frac14\,\psi_{\mu-}\,F_1
          -\frac14\,\gamma_0\psi_{\mu-}\,F_2  +\frac{w}{4}\,\gamma^0\phi_\mu\,\varphi_1  +\frac{w}{4}\,\phi_\mu\,\varphi_2 \,, \\
 \hat D_\mu\chi_- &= D_\mu\chi_- -w\,b_\mu\,\chi_- -\big(\frac{2w}{3}+1\big)\,r_\mu\,\gamma_0\chi_-  \\
     &\quad -\frac14\,\gamma^a\psi_{\mu+}\,e^\rho{}_a\hat D_\rho\varphi_1
          +\frac14\,\gamma^{a0}\psi_{\mu+}\,e^\rho{}_a\hat D_\rho\varphi_2  -\frac14\,\psi_{\mu+}\,F_1
          +\frac14\,\gamma_0\psi_{\mu+}\,F_2 \,.
\end{split}\end{align}
This completes our derivation of the non-relativistic scalar multiplet. In section \ref{sec:TSNC} we will use this scalar
multiplet to derive a new off-shell formulation of Newton--Cartan supergravity.

\subsection{The vector multiplet \label{subsec:vector}}

The $\mathcal{N}=2$ vector multiplet in three dimensions contains a vector, a physical scalar, two spinors and an auxiliary scalar
$(C_\mu,\rho,\lambda_i,D)$. Using the three-dimensional epsilon symbol we can define a new ``dual'' vector
$V_\mu=\varepsilon_\mu{}^{\nu\rho}\,\partial_\nu C_\rho$  with 
\begin{align}\label{vconstr}
 \partial^\mu V_\mu =0 \,,
\end{align}
which has the dimension of an auxiliary field. In terms of  $(\rho,\lambda_i,V_\mu,D)$ we have the following transformation rules:
\begin{align}\begin{split}\label{rigidrelvector2}
 \delta \rho &= \varepsilon^{ij}\,\bar\eta_i\lambda_j \,, \\
 \delta \lambda_i &= -\frac12\,\gamma^\mu \eta_i\,V_\mu -\frac12\,\varepsilon^{ij}\eta_j\,D
              -\frac14\,\gamma^\mu \varepsilon^{ij}\eta_j\,\partial_\mu\rho \,, \\
 \delta D &= \frac12\,\varepsilon^{ij}\,\bar\eta_i\,\gamma^\mu \partial_\mu\lambda_j \,, \\
 \delta V_\mu &= \frac12\,\delta^{ij}\,\bar\eta_i\,\gamma_\mu{}^\nu\partial_\nu\lambda_j \,.
\end{split}\end{align}

Next, we perform the non-relativistic limiting procedure. First, we have to find the scalings of the fields, starting with
the scalings of the supersymmetry parameters given in eq.~\eqref{espscale}. We define new spinors
\begin{align}
 \lambda_\pm = \frac{\omega^{-1\pm1/2}}{\sqrt 2}\,(\lambda_1 \pm\gamma_0\lambda_2) \,,
\end{align}
and the bosonic field
\begin{align}
 \phi = \frac{\rho}{\omega} \,.
\end{align}
Furthermore, we find it useful to introduce the new fields
\begin{align}\begin{split}\label{bosdef}
 S = -\frac1\omega\,V_0 -D \,, \hskip1cm F = \frac{1}{\omega^3}\,V_0 -\frac{1}{\omega^2}\,D\,,
    \hskip1cm C_i=\frac1\omega\,\big( V_i +\frac12\,\varepsilon^{ij}\,\partial_j\rho\big)\,.
\end{split}\end{align}
In the limit $\omega\to\infty$ this leads to the following supersymmetry transformations:
\begin{align}\begin{split}\label{firstvector}
 \delta \phi &= \bar\epsilon_+\gamma^0\lambda_+ -\bar\epsilon_-\gamma^0\lambda_- \,, \\
 \delta \lambda_+ &= \frac14\,\epsilon_+\,\partial_t\phi -\frac12\,\gamma_0\epsilon_+\,S
                     +\frac12\,\gamma^{i0}\epsilon_-\,\partial_i\phi
                     -\frac12\,\gamma^i\epsilon_-\,C_i\,, \\
 \delta S &= \frac12\,\bar\epsilon_+\,\partial_t\lambda_+ -\bar\epsilon_-\gamma^{i0}\partial_i\lambda_+
            -\frac12\,\bar\epsilon_-\,\partial_t\lambda_- \,, \\
 \delta C_i &= \bar\epsilon_-\gamma^{ij}\partial_j\lambda_- +\frac12\,\bar\epsilon_+\gamma^{i0}\partial_t\lambda_- \,, \\
 \delta \lambda_- &= -\frac12\,\gamma^i\epsilon_+\,C_i
            +\frac12\,\gamma_0\epsilon_-\,F \,, \\[.1truecm]
 \delta F &= \bar\epsilon_+\gamma^{i0}\partial_i\lambda_- \,.
\end{split}\end{align}
To prove closure one has to use the constraint
\begin{align}
 \partial^iC_i = \frac12\,\partial_tF \,,
\end{align}
which follows from inserting the definitions \eqref{bosdef} in the relativistic constraint \eqref{vconstr} and sending
$\omega\to\infty$.

An effect of taking the non-relativistic limit is that there exists a consistent truncation of this multiplet. We can impose
\begin{align}\label{limitconstr}
 C_i =0 \,, \hskip2cm F=0 \,, \hskip2cm \lambda_-=0 \,,
\end{align}
which results into
\begin{align}\begin{split}\label{rigidvector}
 \delta \phi &= \bar\epsilon_+\gamma^0\lambda_+ \,, \\
 \delta \lambda_+ &= \frac14\,\epsilon_+\,\partial_t\phi -\frac12\,\gamma_0\epsilon_+\,S
                     +\frac12\,\gamma^{i0}\epsilon_-\,\partial_i\phi \,, \\
 \delta S &= \frac12\,\bar\epsilon_+\,\partial_t\lambda_+ -\bar\epsilon_-\gamma^{i0}\partial_i\lambda_+ \,.
\end{split}\end{align}
While this multiplet looks like a scalar multiplet and appears to be simpler than the scalar multiplet given in \eqref{rigidscalar}
its relation to the relativistic vector multiplet manifests itself in the following way. Due to the redefinition \eqref{bosdef} the
auxiliary field $S$ is related to the zero component of the vector field. As a consequence of this the auxiliary field transforms
non-trivially under Galilean boosts. This can already be seen in the rigid transformations but we will only give the bosonic
transformations when we couple \eqref{rigidvector} to Schr\"odinger supergravity.

After coupling to supergravity the bosonic transformations read
\begin{align}\begin{split}\label{matterbos}
 \delta \phi &= w\,\Lambda_D\phi \,, \\[.1truecm]
 \delta \lambda &= \frac14\,\lambda^{ab}\gamma_{ab}\lambda +(w-1)\,\Lambda_D\,\lambda -\rho\,\gamma_0\lambda \,,\\
 \delta S &= (w-2)\,\Lambda_D\,S-\frac12\,\varepsilon^{ab}\lambda^a\,e^\mu{}_b\hat D_\mu\phi\,,
\end{split}\end{align}
while the fermionic ones take the form
\begin{align}\begin{split}\label{matterferm}
 \delta \phi &= \bar\epsilon_+\gamma^0\lambda \,, \\[.1truecm]
 \delta \lambda &= \frac14\,\epsilon_+\,\tau^\mu\hat D_\mu\phi +\frac12\,\gamma^{a0}\epsilon_-\,e^\mu{}_a\hat D_\mu\phi
                  -\frac12\,\gamma_0\epsilon_+\,S -\frac{w}4\,\eta\,\phi \,,\\
 \delta S &= \frac12\,\bar\epsilon_+\tau^\mu\hat D_\mu\lambda -\bar\epsilon_-\gamma^{a0} e^\mu{}_a\hat D_\mu\lambda
             -\frac{w-1}2\,\bar\eta\,\lambda\,.
\end{split}\end{align}
Note the non-trivial transformation of $S$ under local Galilean boosts, see eq.~\eqref{matterbos}. This makes clear the vector
multiplet origin of \eqref{matterbos} and \eqref{matterferm}. In the formulas above we use the covariant derivatives
\begin{align}\begin{split}
 \hat D_\mu \phi &= \partial_\mu \phi-\bar\psi_{\mu+}\gamma^0\lambda -w\,b_\mu\,\phi \,, \\
 \hat D_\mu \lambda &= \partial_\mu\lambda  -\frac14\,\omega_\mu{}^{ab}\gamma_{ab}\lambda  -(w-1)\,b_\mu\lambda
             +r_\mu\,\gamma_0\lambda   +\frac{w}4\,\phi_\mu\,\phi \\
     &\quad  -\frac14\,\psi_{\mu+}\,\tau^\nu\hat D_\nu\phi  -\frac12\,\gamma^{a0}\psi_{\mu-}\,e^\nu{}_a\hat D_\nu\phi
             +\frac12\,\gamma_0\psi_{\mu+}\,S  \,, \\
 \hat D_\mu S &= \partial_\mu S +2\,b_\mu\,S -\frac12\,\bar\psi_{\mu+}\tau^\rho\hat D_\rho\lambda
             +\bar\psi_{\mu-}\gamma^{a0}e^\rho{}_a\hat D_\rho\lambda \\
     &\quad +\frac12\,\varepsilon^{ab}\,\omega_\mu{}^a\,e^\rho{}_b\hat D_\rho\phi +\frac{w-1}2\,\bar\lambda\,\phi_\mu \,.
\end{split}\end{align}

This finishes our derivation of the non-relativistic vector multiplet. In the following section we will use the non-relativistic
scalar and vector multiplets to derive two inequivalent off-shell formulations of Newton--Cartan supergravity with torsion. Before
doing so we will give a brief overview of the multiplets that we have discussed so far and which provide the basis of a
non-relativistic superconformal tensor calculus, see table \ref{tab:1}.
\begin{table}[ht]
\begin{center}
{\bf Overview of non-relativistic multiplets} \\[2mm]
{\small
\begin{tabular}{|c||c|c|c|c|}
\hline \rule[-1mm]{0mm}{6mm}
 multiplet & field & type & $D$-weight & $R$-weight \\[.1truecm]
\hline \rule[-1mm]{0mm}{6mm}

Schr\"odinger & $\tau_\mu$ & time-like vielbein & 2 & 0 \\
             & $e_\mu{}^a$ & spatial vielbein & 1 & 0 \\
              & $m_\mu$    & $Z$ gauge field & 0 & 0 \\
              & $r_\mu$    & $R$ gauge field & 0 & 0 \\
              & $b$        & ``$D$ gauge field'' & -2 & 0 \\
              & $\psi_{\mu+}$ & $Q_+$ gravitino & 1 & -1 \\
              & $\psi_{\mu-}$ & $Q_-$ gravitino & 0 & 1 \\[.1truecm]
\hline \rule[-1mm]{0mm}{6mm}

Scalar & $\varphi_1$ & physical scalar & $w$ & $\frac{2w}{3}$ \\
       & $\varphi_2$ & physical scalar & $w$ & $-\frac{2w}{3}$ \\
       & $\chi_+$    & spinor          & $w-1$ & $-\frac{2w}{3}-1$ \\
       & $\chi_-$    & spinor          & $w$   & $\frac{2w}{3}+1$ \\
       & $F_1$       & auxiliary scalar & $w-1$ & $\frac{2w}{3} +2$ \\
       & $F_2$       & auxiliary scalar & $w-1$ & $-\frac{2w}{3} -2$ \\[.1truecm]
\hline \rule[-1mm]{0mm}{6mm}

Vector & $\phi$ & physical scalar & $w$ & 0 \\
       & $\lambda$ & spinor       & $w-1$ & -1 \\
       & $S$    & auxiliary       & $w-2$ & 0 \\[.1truecm]
\hline

\end{tabular}
}
\end{center}

 \caption{\sl Properties of three-dimensional non-relativistic multiplets.
         }\label{tab:1}
\end{table}

Note that if we were to add another column to this table for the central charge weight ($Z$-weight) we would have only zeros.
We will come back to this in the conclusion section.


\section{Newton--Cartan supergravity with torsion \label{sec:TSNC}}

At this point we have at our disposal a ``conformal'' Schr\"odinger supergravity theory and two matter multiplets which we can use
to fix some of the gauge symmetries. This enables us to use superconformal techniques to derive off-shell non-relativistic
supergravity multiplets. The superconformal tensor calculus naturally leads to a Newton--Cartan supergravity with non-zero torsion,
i.e.~the curl of the gauge field $\tau_\mu$ of local time translations is non-zero, see also \cite{Bergshoeff:2014uea} for a
discussion of the bosonic case. The origin of the torsion is the spatial part $b_a$ of the dilatation gauge field. Unlike in the
relativistic case, this spatial part cannot be shifted away by a special conformal transformation. Instead, it is a dependent gauge
field whose presence leads to torsion.

In this section we show how the extra symmetries of the Schr\"odinger superalgebra that are not contained in the Bargmann
superalgebra, i.e.~dilatations $D$, special conformal transformations $K$, $S$-supersymmetry and possibly $R$-symmetry, can be
eliminated by using a compensator matter multiplet. First, we eliminate the special conformal transformations by setting
\begin{align}\label{nob}
 b = \tau^\mu b_\mu =0 \,.
\end{align}
The induced compensating transformation is given in eq.~\eqref{lambdaKcomp}. This step is the same independent of which compensator
multiplet we use. In the following we shall use both, the scalar and the vector multiplet from the previous section. In analogy to
the relativistic case we refer to the resulting off-shell formulations as the ``old minimal'' one when we use a compensator scalar
multiplet and the ``new minimal'' formulation when the compensator multiplet is the vector multiplet.

\subsection{The ``old minimal'' formulation \label{subsec:old}}

In this subsection we choose  the scalar multiplet whose transformation rules can be found in eqs.~\eqref{schroscalarbos} and
\eqref{schroscalarferm} as the compensator multiplet. Like in the relativistic case we eliminate both physical scalars thus gauge
fixing the dilatations and  the local $U(1)$ $R$-symmetry. One of the fermions is used to get rid of the special conformal
$S$-supersymmetry:
\begin{align}
 \begin{aligned}  \varphi_1 =1 \,: \\ \varphi_2 =0 \,: \end{aligned}\quad\bigg\}
                \hskip1.33cm  &\text{fixes dilatations and $R$-symmetry} \,, \label{fix1} \\
 \chi_+ =0 \,: \hskip2cm  &\text{fixes special conformal $S$-supersymmetry} \,.
\end{align}
The compensating transformations are given by
\begin{align}\label{Dandrhocomp}
 \Lambda_D &= -\frac1w\,\bar\epsilon_-\chi_- \,, \hskip2cm    \rho = -\frac{3}{2w}\,\bar\epsilon_-\gamma^0\chi_- \,,
\end{align}
and
\begin{align}\begin{split}\label{easyetacomp}
 \eta &= -\frac1w\,\epsilon_+\,\tau^\mu\bar\psi_{\mu-}\chi_-
        +\gamma_0\epsilon_+\,\tau^\mu\Big(\frac23\,r_\mu +\frac1w\,\bar\psi_{\mu-}\gamma^0\chi_-\Big)
        -\gamma^{a0}\epsilon_-\Big(b_a +\frac1w\,e^\mu{}_a\,\bar\psi_{\mu-}\chi_-\Big) \\
     &\quad -\gamma^a\epsilon_-\,e^\mu{}_a\,\Big(\frac23\,r_\mu +\frac1w\,\bar\psi_{\mu-}\gamma^0\chi_-\Big)
        +\frac1w\,\gamma_0\epsilon_-\,F_1  -\frac1w\,\epsilon_-\,F_2  -\frac2w\,\lambda^a\gamma_a\chi_- \,.
\end{split}\end{align}
We thus end up with the field content  given in eq.~\eqref{om} of the ``old minimal'' Newton--Cartan supergravity theory that
realizes the Bargmann superalgebra off-shell.
The transformation rules of all fields can be easily constructed using  those of Schr\"odinger supergravity, see section
\ref{sec:schroedinger}, and those of the scalar multiplet, see eqs.~\eqref{schroscalarbos} and \eqref{schroscalarferm}, together with
the compensating transformations given in eqs.~\eqref{lambdaKcomp}, \eqref{Dandrhocomp} and \eqref{easyetacomp}. Given the lengthy
nature of the final transformation rules we have moved the explicit expressions to appendix \ref{app:dependentfields}.

\subsection{The ``new minimal'' formulation \label{subsec:new}}

In this subsection we choose the vector multiplet, see eqs.~\eqref{matterbos} and \eqref{matterferm}, as the compensator multiplet.
The gauge fixing of dilatations and the special conformal $S$-supersymmetry is done by imposing the conditions
\begin{align}\begin{aligned}
 \phi &=1 \,: \hskip1cm & & \textrm{fixes dilatations}\,, \\
 \lambda &=0 \,: & & \textrm{fixes }S\textrm{-supersymmetry} \,,
\end{aligned}\end{align}
and the resulting compensating gauge transformations are given by
\begin{align}\label{lDetacomp}
 \Lambda_D =0 \,, \hskip2cm \eta= -\frac2w\,\gamma_0\epsilon_+\,S -2\,\gamma^{a0}\epsilon_-\,b_a \,.
\end{align}
At this point we are left with the symmetries of the Bargmann superalgebra, see eqs.~\eqref{Bargmannalg} and
\eqref{3dsuperbargmann}, plus an extra $U(1)$ $R$-symmetry. These symmetries  are realized on the  set of independent fields
of the ``new minimal'' Newton--Cartan supergravity theory given in eq.~\eqref{nm}.
This theory is the non-relativistic version of the three-dimensional $\mathcal{N}=(2,0)$ new minimal
Poincar\'e supergravity theory. The bosonic transformations of the different fields are given by
\begin{align}\begin{split}
 \delta \tau_\mu &=0 \,, \\
 \delta e_\mu{}^a &= \lambda^a{}_b\,e_\mu{}^b +\tau_\mu\,\lambda^a \,, \\
 \delta m_\mu &= \partial_\mu\sigma +\lambda^ae_\mu{}^a \,, \\
 \delta r_\mu &= \partial_\mu\rho \,, \\
 \delta S &= -\frac12\,\varepsilon^{ab}\lambda^ab_b \,,
\end{split}\end{align}
and
\begin{align}\begin{split}
 \delta \psi_{\mu+} &= \frac14\,\lambda^{ab}\gamma_{ab}\psi_{\mu+} -\gamma_0\psi_{\mu+}\,\rho \,, \\
 \delta \psi_{\mu-} &= \frac14\,\lambda^{ab}\gamma_{ab}\psi_{\mu-} -\frac12\,\lambda^a\gamma_{a0}\psi_{\mu+}
                      +\gamma_0\psi_{\mu+}\,\rho \,.
\end{split}\end{align}
Note that $S$ transforms non-trivially under a Galilean  boost transformation which is proportional to $b_a$, i.e.~to torsion, see
eq.~\eqref{torsion}. The
fermionic transformations including the compensating terms that follow from eq.~\eqref{lDetacomp} are given by
\begin{align}\begin{split}
 \delta \tau_\mu &= \frac12\,\bar\epsilon_+\gamma^0\psi_{\mu+} \,, \\
 \delta e_\mu{}^a &= \frac12\,\bar\epsilon_+\gamma^a\psi_{\mu-} +\frac12\,\bar\epsilon_-\gamma^a\psi_{\mu+} \,,\\
 \delta m_\mu &= \bar\epsilon_-\gamma^0\psi_{\mu-} \,, \\
 \delta r_\mu &= \frac34\,\bar\epsilon_-\gamma^{a0}\psi_{\mu+}\,b_a -\frac38\,\bar\epsilon_+\phi_\mu
               -\frac3{4w}\,\bar\epsilon_+\gamma^0\psi_{\mu+}\,S \,, \\
 \delta S &= \frac{w}8\,\bar\epsilon_+\phi_\mu\,\tau^\mu
           +\frac{w}4\,\bar\epsilon_+\gamma^{a0}\psi_{\mu-}\,\tau^\mu\,b_a -\frac14\,\bar\epsilon_+\gamma^0\psi_{\mu+}\,S \\
   &\quad  -\frac{w}4\,\bar\epsilon_-\gamma^{a0}\phi_\mu\,e^\mu{}_a
           -\frac{w}2\,\bar\epsilon_-\gamma^a\gamma^b\psi_{\mu-}\,e^\mu{}_a\,b_b
           -\frac12\,\bar\epsilon_-\gamma^a\psi_{\mu+}\,e^\mu{}_a\,S \,,
\end{split}\end{align}
and
\begin{align}\begin{split}
 \delta \psi_{\mu+} &= D_\mu\epsilon_+ -\epsilon_+\,e_\mu{}^a\,b_a +\gamma_0\epsilon_+\,r_\mu
               +\frac2w\,\gamma_0\epsilon_+\,\tau_\mu\,S +2\,\gamma^{a0}\epsilon_-\,\tau_\mu\,b_a \,, \\
 \delta \psi_{\mu-} &= D_\mu\epsilon_- -\gamma_0\epsilon_-\,r_\mu +\gamma^a\gamma^b\epsilon_-\,e_\mu{}^a\,b_b
               +\frac12\,\gamma_{a0}\epsilon_+\,\omega_\mu{}^a
               +\frac1w\gamma_a\epsilon_+\,e_\mu{}^a\,S \,.
\end{split}\end{align}
The transformation rules of the dependent gauge fields can be found in  appendix \ref{app:dependentfields}.


\section{Truncation to zero torsion \label{sec:notorsion}}

In the previous section we derived  a Newton--Cartan supergravity theory with non-zero torsion. This needs to be contrasted with the
Newton--Cartan supergravity theories constructed in \cite{Andringa:2013mma,Bergshoeff:2015uaa} that  have zero torsion. To see the
difference, it is instructive to compare the curvature of local time translations for the theories with and without torsion.
Indicating  the curvature of the torsionfull theory with $\mathcal R(H)$ and the one of the zero-torsion theory with ${\hat R}(H)$
we have
\begin{align}\begin{split}
\mathcal{R}_{\mu\nu}(H) &= 2\,\partial_{[\mu}\tau_{\nu]} -4\,b_{[\mu}\tau_{\nu]}
          -\frac12\,\bar\psi_{[\mu+}\gamma^0\psi_{\nu]+}\,,\\
 \hat R_{\mu\nu}(H) &= 2\,\partial_{[\mu}\tau_{\nu]} -\frac12\,\bar\psi_{[\mu+}\gamma^0\psi_{\nu]+}\,.\label{2dlinec}
 \end{split}\end{align}
Note that the space-space components of both curvatures are the same. The difference is in the time-space component. In the
torsionfull case, setting the time-space component to zero, is a conventional constraint that is used to solve for the spatial
part $b_a$ of the dilatation gauge field whereas in the torsionless case it represents an un-conventional constraint. Indeed, we have 
\begin{align}\label{torsion}
 b_a = \frac12\,\hat R_{a0}(H) \,,
\end{align}
and therefore setting the torsion to zero, i.e.
\begin{equation}\label{zerotorsion}
b_a=0\,, 
\end{equation}
leads to the un-conventional constraint $\hat R_{a0}(H)$ in the torsionless theory.

This points us to an interesting observation: the existence of a non-trivial truncation of
the old minimal and new minimal Newton--Cartan supergravity multiplets constructed in  section \ref{sec:TSNC}. Indeed, we shall show
in this section how we can reduce the new minimal torsionfull theory constructed in subsection \ref{subsec:new}
to the known new minimal torsionless Newton--Cartan supergravity theory
constructed in \cite{Andringa:2013mma,Bergshoeff:2015uaa}.

We now investigate the consequences of imposing the zero-torsion constraint \eqref{zerotorsion}.  It is convenient to use the
explicit expression for the $S$-supersymmetry gauge field field $\phi_\mu$, which simplifies to
\begin{align}
 \phi_\mu &= \gamma^{a0}\hat\psi_{a\mu-} -\frac2w\,\gamma_0\psi_{\mu+}\,S \,,
\end{align}
when we use the curvatures and constraints that we introduce below. The only dependent gauge fields of the Newton--Cartan supergravity
theory are the connection fields for spatial rotations and Galilean boosts. For the supersymmetry rules of the independent gauge
fields we find
 \begin{align}\begin{split}
 \delta \tau_\mu &= \frac12\,\bar\epsilon_+\gamma^0\psi_{\mu+} \,, \\
 \delta e_\mu{}^a &= \frac12\,\bar\epsilon_+\gamma^a\psi_{\mu-} +\frac12\,\bar\epsilon_-\gamma^a\psi_{\mu+} \,,\\
 \delta m_\mu &= \bar\epsilon_-\gamma^0\psi_{\mu-} \,, \\
 \delta r_\mu &= -\frac38\,\bar\epsilon_+\gamma^{a0}\hat\psi_{a\mu-} -\frac3{2w}\,\bar\epsilon_+\gamma^0\psi_{\mu+}\,S \,, \\
 \delta S &= \frac{w}8\,\bar\epsilon_+\gamma^{a0}\hat\psi_{a0-} \,,
\end{split}\end{align}
and
\begin{align}\begin{split}
 \delta \psi_{\mu+} &= D_\mu\epsilon_+ +\gamma_0\epsilon_+\,r_\mu +\frac2w\,\gamma_0\epsilon_+\,\tau_\mu\,S \,, \\
 \delta \psi_{\mu-} &= D_\mu\epsilon_- -\gamma_0\epsilon_-\,r_\mu +\frac12\,\gamma_{a0}\epsilon_+\,\omega_\mu{}^a
               +\frac1w\,\gamma_a\epsilon_+\,e_\mu{}^a\,S \,.
\end{split}\end{align}
The  curvatures and derivatives of the new minimal torsionless Newton--Cartan supergravity theory are now given by \eqref{2dlinec} and
\begin{align}\begin{split}
 \hat R_{\mu\nu}{}^a(P) &= 2\,\partial_{[\mu}e_{\nu]}{}^a -2\,\omega_{[\mu}{}^{ab}e_{\nu]}{}^b -2\,\omega_{[\mu}{}^a\tau_{\nu]}
               -\bar\psi_{[\mu+}\gamma^a\psi_{\nu]-} \,, \\[.15truecm]
 \hat R_{\mu\nu}(Z) &= 2\,\partial_{[\mu}m_{\nu]} -\bar\psi_{[\mu-}\gamma^0\psi_{\nu]-}  \,, \\
 \hat R_{\mu\nu}(R) &= 2\,\partial_{[\mu}r_{\nu]} +\frac3{2w}\,\bar\psi_{[\mu+}\gamma^0\psi_{\nu]+}\,S
               +\frac34\,\bar\psi_{[\mu+}\gamma^{a0}\hat\psi_{a\nu]-}  \,, \\
 \hat D_\mu S &= \partial_\mu S -\frac{w}8\,\bar\psi_{\mu+}\gamma^{a0}\hat\psi_{a0-} \,, \\
 \hat \psi_{\mu\nu+} &= 2\,\partial_{[\mu}\psi_{\nu]+} -\frac12\,\omega_{[\mu}{}^{ab}\gamma_{ab}\psi_{\nu]+}
               -2\,\gamma_0\psi_{[\mu+}\,r_{\nu]} -\frac4w\,\gamma_0\psi_{[\mu+}\,\tau_{\nu]}\,S \,, \\
 \hat \psi_{\mu\nu-} &= 2\,\partial_{[\mu}\psi_{\nu]-} -\frac12\,\omega_{[\mu}{}^{ab}\gamma_{ab}\psi_{\nu]-}
               +2\,\gamma_0\psi_{[\mu-}\,r_{\nu]} +\omega_{[\mu}{}^a\gamma_{a0}\psi_{\nu]+}
               -\frac2w\,\gamma_a\psi_{[\mu+}\,e_{\nu]}{}^a\,S \,.
\end{split}\end{align}

As we explained at the beginning of this section, the zero-torsion constraint \eqref{zerotorsion} may convert a conventional
constraint into an un-conventional one.
If this happens we have to check if the supersymmetry variation of this un-conventional constraint leads to further constraints. To
perform this check we need the transformation rules of the dependent connection gauge fields which reduce to
\begin{align}\begin{split}
 \delta \omega_\mu{}^{ab} &= -\frac12\,\bar\epsilon_+\gamma^{[a}\hat\psi^{b]}{}_{\mu-}
               +\frac1w\,\bar\epsilon_+\gamma^{ab}\psi_{\mu+}\,S \,, \\
 \delta \omega_\mu{}^a &= \bar\epsilon_-\gamma^0\hat\psi_\mu{}^a{}_- +\frac14\,e_\mu{}^b\,\bar\epsilon_+\gamma^b\hat\psi^a{}_{0-}
              +\frac14\,\bar\epsilon_+\gamma^a\hat\psi_{\mu0-}  -\frac1w\,\bar\epsilon_+\gamma^{a0}\psi_{\mu-}\,S
              -\frac1w\,\bar\epsilon_-\gamma^{a0}\psi_{\mu+}\,S  \,.
\end{split}\end{align}
The corresponding curvatures are given by
\begin{align}\begin{split}
 \hat R_{\mu\nu}{}^{ab}(J) &= 2\,\partial_{[\mu}\omega_{\nu]}{}^{ab} +\bar\psi_{[\mu+}\gamma^{[a}\hat\psi^{b]}{}_{\nu]-}
              -\frac1w\,\bar\psi_{[\mu+}\gamma^{ab}\psi_{\nu]+}\,S  \,, \\
 \hat R_{\mu\nu}{}^a(G) &= 2\,\partial_{[\mu}\omega_{\nu]}{}^a -2\,\omega_{[\mu}{}^{ab}\omega_{\nu]}{}^b
              -2\,\bar\psi_{[\mu-}\gamma^0\hat\psi_{\nu]}{}^a{}_- -\frac12\,e_{[\nu}{}^b\bar\psi_{\mu]+}\gamma^b\hat\psi^a{}_{0-} \\
     &\quad   -\frac12\,\bar\psi_{[\mu+}\gamma^a\hat\psi_{\nu]0-} +\frac2w\,\bar\psi_{[\mu+}\gamma^{a0}\psi_{\nu]-}\,S \,.
\end{split}\end{align}

We are now ready to discuss the constraint structure of the truncated theory. Some of the curvatures did not change, hence we can
immediately infer, e.g., that
\begin{align}\label{oldconstr}
 \hat R_{ab}(R) =0 \,, \hskip2cm \frac34\,\varepsilon^{ab}\,\hat R_{\mu\nu}{}^{ab}(J)=\hat R_{\mu\nu}(R) \,.
\end{align}
The constraints $\hat R_{\mu\nu}{}^a(P)=0$ and $\hat R_{\mu\nu}(Z)=0$ are identities when we insert the expressions for the connection
gauge fields, i.e.~they are conventional constraints. More importantly though, we find new constraints. This is due to the
fact that we imposed $\hat R_{a0}(H)=0$ which is an example of a conventional constraint (necessary to solve for the spatial part
$b_a$ of the dilatation gauge field) that gets converted into an un-conventional constraint. Together with the constraint 
$\hat R_{ab}(H)=0$ which reads the same in the torsionfull as well as in the torsionless case, we find $\hat R_{\mu\nu}(H)=0$.
Supersymmetry variations of this constraint reveal the following additional constraints:
\begin{align}
  \stackrel{Q_-}{\longrightarrow} \hskip1.1cm \hat\psi_{ab-} &=0 \label{offmoreconstraints}\\
 \hat R_{\mu\nu}(H)=0 \quad \stackrel{Q_+}{\longrightarrow} \quad \hat\psi_{\mu\nu+}=0 \quad
     \stackrel{Q_+}{\longrightarrow} \quad
 \hat R_{\mu\nu}{}^{ab}(J) &=\frac4w\,\varepsilon^{ab}\,\tau_{[\mu}\hat D_{\nu]}\,S \,. \label{offconstraints}
\end{align}
Further transformations only lead to Bianchi identities. By combining the constraints  \eqref{offconstraints} with \eqref{oldconstr}
we furthermore derive that 
\begin{align}
 -\frac6w\,\hat D_{[\mu}\big(\tau_{\nu]}\,S\big) = 2\,\hat D_{[\mu}r_{\nu]} \,.
\end{align}
This constraint implies that up to an arbitrary constant the $R$-symmetry gauge field $r_\mu$ is determined by $\tau_\mu$ and $S$.
In fact, when we set
\begin{align}\label{nor}
 r_\mu = -\frac3w\,\tau_\mu\,S \,,
\end{align}
the truncated theory leads to the off-shell Newton--Cartan multiplet that was presented in \cite{Bergshoeff:2015uaa}.
Furthermore, by making the redefinition
\begin{align}
 r_\mu=-V_\mu -\frac1w\,\tau_\mu\,S \,,
\end{align}
one obtains precisely the off-shell multiplet that is obtained when taking the limit of the
new minimal Poincar\'e multiplet as described in \cite{Bergshoeff:2015uaa}.


\section{Conclusions and outlook \label{sec:conclusions}}

In this  paper we have discussed extensions of non-relativistic supergravity to include conformal symmetries. As an example we
have constructed a three-dimensional theory of Schr\"odinger supergravity, i.e.~a theory that realizes a Schr\"odinger
superalgebra, and we have successfully constructed two matter multiplets. These results are summarized in table \ref{tab:1}. We
have then introduced a non-relativistic version of the superconformal tensor calculus and used it to construct two inequivalent
off-shell formulations, called the old minimal and new minimal formulation, of a three-dimensional non-relativistic Newton--Cartan
supergravity multiplet with torsion.

The appearance of torsion is one of the points where our analysis differs from the relativistic one. In the relativistic case the full gauge field of
dilatations $b_\mu$ is a St\"uckelberg field for special conformal transformations and the theory is by construction torsionless.
In contrast, in the non-relativistic case only the time component $b$ is a St\"uckelberg field for the single (scalar) special
conformal transformation of the Schr\"odinger superalgebra. The spatial components $b_a$ on the other hand are dependent gauge
field components and they are proportional to torsion. Thus, unless we set set $b_a=0$ as we did in section \ref{sec:notorsion},
the superconformal approach always leads to torsionfull theories in the non-relativistic setting.

It would be interesting to see how one can go on-shell in the presence of torsion. This is not a straightforward thing to do since
to our knowledge even in the bosonic case the equations of motion describing Newton--Cartan gravity with torsion have not been
written down so far.\,\footnote{
A systematic approach to construct such an equation of motion will be given in \cite{Blaisepaper}.}
Even in the absence of torsion the equations of motion have only been written down under the assumption that the curvature of
spatial rotations is zero \cite{Andringa:2010it}. It is not difficult to write down the equations of motion for the case that
this curvature is nonzero but the price one has to pay is that one has to add extra terms to the equation of motion proposed in
\cite{Andringa:2010it} that break the invariance under central charge transformations \cite{Bergshoeff:2014uea}. In the bosonic
case this extended equation of motion can be understood by applying a conformal tensor calculus at the level of the equations
of motion (without the need to write down an action) using a single compensator scalar transforming under dilatations. The
situation gets more intricate when one introduces non-zero torsion because in that case a second compensating scalar is needed
that transforms non-trivially under central charge transformations. This second compensating scalar should therefore be part of
a different multiplet than the scalar and vector multiplets we considered in this work. The construction of such a multiplet is
different from our investigations in section \ref{sec:matter} and goes beyond the scope of this paper. We hope to return to the
issue of how to go on-shell with a non-flat foliation space and in the presence of torsion in a future work.

Perhaps we can get some inspiration from a similar problem in the relativistic case. In the four-dimensional $\mathcal{N}=2$
off-shell formulation one also has to use two compensator multiplets in order to be able to write down an action \cite{deWit:1982na}.
The first compensator multiplet fixes dilatations, $S$-supersymmetry and a chiral $U(1)$ symmetry. The second compensator multiplet
fixes a remaining local chiral $SU(2)$ symmetry and it is needed only to be able to write down an action. In our analogy this would
correspond to fixing central charge symmetry. Maybe a non-relativistic matter multiplet with a scalar field that has a non-trivial
central charge transformation could be found as a non-relativistic (three-dimensional) analogue of one of the three compensator
multiplets used in \cite{deWit:1982na}.

In this paper we only considered the construction of pure Newton--Cartan supergravity. A natural generalization of our work would be
to consider general non-relativistic matter-coupled Newton--Cartan supergravity theories with simple or extended supersymmetry. This
would answer the question of what the non-relativistic analogue is of the geometries that one encounters in the relativistic
matter-coupled supergravity theories. For example, it would be interesting to find out what the non-relativistic analogue is of a
K\"ahler target space.

Finally, it would be very interesting to find higher-dimensional analogues of our results on Newton--Cartan supergravity. So far,
the use of gauging techniques has failed to lead to e.g.~a four-dimensional theory of Newton--Cartan supergravity. 
It is a priori not clear what auxiliary fields are needed to close the supersymmetry algebra. Presumably,
similar obstacles are encountered if one were to try to gauge a four-dimensional Schr\"odinger superalgebra. The limiting
procedure discussed in \cite{Bergshoeff:2015uaa}, if its application is equally straightforward in higher dimensions, might be the
simplest way to find  a four-dimensional Newton--Cartan supergravity theory.


\section*{Acknowledgements}

We are grateful to Hamid Afshar, Joaquim Gomis and Blaise Rollier for discussions.
JR was supported by the START project Y 435-N16 of the Austrian Science Fund (FWF) and by the NCCR SwissMAP, funded by the Swiss
National Science Foundation.
TZ acknowledges financial support by the Dutch Academy of Sciences (KNAW).


\appendix


\section{Details on the off-shell multiplets \label{app:dependentfields}}

This appendix contains more details about the two off-shell formulations of torsional Newton--Cartan supergravity that
feature in the main text. In particular, we give the transformation rules of all independent fields of the old minimal formulation
in appendix \ref{app:old}. Those for the new minimal formulation were given in section \ref{subsec:new}. In appendix \ref{app:new} we
give the transformation rules of the dependent gauge fields of the new minimal formulation which are needed to show that the
commutator algebra closes.

\subsection{``Old minimal'' formulation \label{app:old}}

We collect  here the transformation rules of the independent gauge fields of the  old minimal formulation.  We find that the bosonic
gauge fields transform as follows under the bosonic transformations
\begin{align}\begin{split}\label{oldbos}
 \delta \tau_\mu &=0 \,, \\
 \delta e_\mu{}^a &= \lambda^a{}_b\,e_\mu{}^b +\tau_\mu\,\lambda^a \,, \\
 \delta m_\mu &= \partial_\mu\sigma +\lambda^ae_\mu{}^a \,, \\
 \delta r_\mu &= -\frac{3}{4w}\,\lambda^a\,\bar\psi_{\mu+}\gamma^a\chi_-  \,, \\
 \delta F_1 &= 0 \,, \\
 \delta F_2 &= 0 \,,
\end{split}\end{align}
while the fermionic gauge fields transform as
\begin{align}\begin{split}
 \delta \psi_{\mu+} &= \frac14\,\lambda^{ab}\gamma_{ab}\psi_{\mu+}    +\frac2w\,\tau_\mu\,\lambda^a\gamma_a\chi_- \,, \\
 \delta \psi_{\mu-} &= \frac14\,\lambda^{ab}\gamma_{ab}\psi_{\mu-} -\frac12\,\lambda^a\gamma_{a0}\psi_{\mu+}
             +\frac1w\,e_\mu{}^a\,\lambda^b\,\gamma_a\gamma_{b0}\chi_- \,, \\
 \delta \chi_- &= \frac14\,\lambda^{ab}\gamma_{ab}\chi_- \,.
\end{split}\end{align}
Note the non-trivial Galilean boost transformation of the $R$-symmetry gauge field $r_\mu$ in \eqref{oldbos}. The supersymmetry
transformations are given by
\begin{align}\begin{split}
 \delta \tau_\mu &= \frac12\,\bar\epsilon_+\gamma^0\psi_{\mu+} \,, \\
 \delta e_\mu{}^a &= \frac12\,\bar\epsilon_+\gamma^a\psi_{\mu-} +\frac12\,\bar\epsilon_-\gamma^a\psi_{\mu+} \,,\\
 \delta m_\mu &= \bar\epsilon_-\gamma^0\psi_{\mu-} \,,
\end{split}\end{align}
and
\begin{align}\begin{split}
 \delta \psi_{\mu+} &= D_\mu\epsilon_+ -e_\mu{}^a\,b_a\,\epsilon_+
          +\big(r_\mu -\frac23\,\tau_\mu\tau^\rho r_\rho\big)\,\gamma_0\epsilon_+
          +\frac23\,\gamma^a\epsilon_-\,\tau_\mu\,e^\rho{}_a\,r_\rho +\gamma^{a0}\epsilon_-\,\tau_\mu\,b_a \\
   &\quad -\frac1w\,\gamma_0\epsilon_-\,\tau_\mu\,F_1   +\frac1w\,\epsilon_-\,\tau_\mu\,F_2
          -\frac1w\,\psi_{\mu+}\,\bar\epsilon_-\chi_- +\frac{3}{2w}\,\gamma_0\psi_{\mu+}\,\bar\epsilon_-\gamma^0\chi_- \\
   &\quad +\frac1w\,\tau_\mu\,\gamma^a\chi_-\,\bar\epsilon_+\gamma^a\psi_{\rho-}\,\tau^\rho
          -\frac1w\,\tau_\mu\,\gamma^{a0}\chi_-\,\bar\epsilon_-\psi_{\rho-}\,e^\rho{}_a
          +\frac1w\,\tau_\mu\,\gamma^a\chi_-\,\bar\epsilon_-\gamma^0\psi_{\rho-}\,e^\rho{}_a \,, \\
 \delta \psi_{\mu-} &= D_\mu\epsilon_- -r_\mu\,\gamma_0\epsilon_- +\frac12\,\omega_\mu{}^a\gamma_{a0}\epsilon_+
          -\frac13\,\gamma^a\epsilon_+\,e_\mu{}^a\tau^\rho\,r_\rho
          -\frac13\,\gamma^a\gamma^{b0}\epsilon_+\,e_\mu{}^ae^\rho{}_b\,r_\rho \\
   &\quad +\frac12\,\gamma^a\gamma^b\epsilon_-\,e_\mu{}^a\,b_b  -\frac1{2w}\,\gamma^a\epsilon_-\,e_\mu{}^a\,F_1
          -\frac1{2w}\,\gamma_{a0}\epsilon_-\,e_\mu{}^a\,F_2  -\frac3{2w}\,\gamma_0\psi_{\mu-}\,\bar\epsilon_-\gamma^0\chi_- \\
   &\quad -\frac{1}{2w}\,\gamma^a\gamma^{b0}\chi_-\,\bar\epsilon_+\gamma^b\psi_{\rho-}\,e_\mu{}^a\,\tau^\rho
          -\frac{1}{2w}\,\gamma^a\gamma^b\chi_-\,\bar\epsilon_-\psi_{\rho-}\,e_\mu{}^ae^\rho{}_b \\
   &\quad -\frac{1}{2w}\,\gamma^a\gamma^{b0}\chi_-\,\bar\epsilon_-\gamma^0\psi_{\rho-}\,e_\mu{}^ae^\rho{}_b \,, \\
 \delta \chi_- &= -\frac{w}{6}\,\gamma^{a0}\epsilon_+\,e^\mu{}_a\,r_\mu   -\frac{w}4\,\gamma^a\epsilon_+\,b_a
          -\frac{1}{3w}\,\epsilon_-\,\bar\chi_-\chi_-  +\frac14\,\epsilon_+\,F_1   -\frac14\,\gamma_0\epsilon_+\,F_2 \\
   &\quad -\frac14\,\gamma^a\gamma^b\chi_- \,\bar\epsilon_+\gamma^b\psi_{\mu-}\,e^\mu{}_a \,.
\end{split}\end{align}
Finally, for the $R$-symmetry gauge field  $r_\mu$ and the auxiliary scalars $F_1$ and $F_2$  we find the following transformations:
\begin{align}\begin{split}
 \delta r_\mu &= -\frac38\,\bar\epsilon_+\phi_\mu  +\frac14\,\bar\epsilon_+\gamma^0\psi_{\mu+}\,\tau^\rho r_\rho
          +\frac14\,\bar\epsilon_-\gamma^{a0}\psi_{\mu+}\,e^\rho{}_ar_\rho +\frac38\,\bar\epsilon_-\gamma^{a0}\psi_{\mu+}\,b_a \\
   &\quad +\frac3{8w}\,\bar\epsilon_-\gamma^0\psi_{\mu+}\,F_1  -\frac3{8w}\,\bar\epsilon_-\psi_{\mu+}\,F_2
          -\frac3{8w}\,\bar\epsilon_+\gamma^a\psi_{\rho-}\,\tau^\rho\,\bar\psi_{\mu+}\gamma^a\chi_- \\
   &\quad +\frac3{8w}\,\bar\epsilon_-\psi_{\rho-}\,e^\rho{}_a\,\bar\psi_{\mu+}\gamma^{a0}\chi_-
          -\frac3{8w}\,\bar\epsilon_-\gamma^0\psi_{\rho-}\,e^\rho{}_a\,\bar\psi_{\mu+}\gamma^a\chi_- \,, \\
 \delta F_1 &= \bar\epsilon_+\gamma^0\tau^\mu\hat D_\mu \chi_- +\bar\epsilon_-\gamma^ae^\mu{}_a\hat D_\mu\chi_-
          +\frac2w\,\bar\epsilon_-\chi_-\,F_1  -\frac2w\,\bar\epsilon_-\gamma^0\chi_- \,F_2  \\
   &\quad -\frac14\,\bar\epsilon_+\gamma^a\psi_{\mu-}\,e^\mu{}_a\,F_1
          +\frac14\,\bar\epsilon_+\gamma^{a0}\psi_{\mu-}\,e^\mu{}_a\,F_2
          +\frac{w}4\,\bar\epsilon_+\gamma^{a0}\phi_\mu\,e^\mu{}_a \\
   &\quad +\frac12\,\bar\epsilon_+\gamma^a\gamma_{b0}\chi_-\,e^\mu{}_a\,\omega_\mu{}^b
          -\frac{w}6\,\bar\epsilon_+\gamma^a\psi_{\mu+}\,e^\mu{}_a\tau^\rho\,r_\rho
          -\frac{w}6\,\bar\epsilon_+\gamma^a\gamma^{b0}\psi_{\mu-}\,e^\mu{}_ae^\rho{}_b\,r_\rho \\
   &\quad +\frac{2(w+1)}{3}\,\bar\epsilon_+\chi_-\,\tau^\mu\,r_\mu
          -\frac{2(w+1)}{3}\,\bar\epsilon_-\gamma^{a0}\chi_-\,e^\mu{}_a\,r_\mu
          +\frac{w}4\,\bar\epsilon_+\gamma^a\gamma^b\psi_{\mu-}\,e^\mu{}_a\,b_b \\
   &\quad +(w+1)\,\bar\epsilon_-\gamma^a\chi_-\,b_a
          +\frac14\,\bar\epsilon_+\gamma^a\gamma^{b0}\chi_-\,\bar\psi_{\rho-}\gamma^b\psi_{\mu+}\,e^\mu{}_a\tau^\rho
          -\frac14\,\bar\epsilon_+\chi_-\,\bar\psi_{\rho-}\psi_{\mu-}\,e^\mu{}_ae^\rho{}_a \\
   &\quad +\frac14\,\bar\epsilon_+\gamma^{ab0}\chi_-\,\bar\psi_{\rho-}\gamma^0\psi_{\mu-}\,e^\mu{}_ae^\rho{}_b \,,
\end{split}\end{align}
and
\begin{align}\begin{split}
 \delta F_2 &= \bar\epsilon_+\gamma^0\tau^\mu\hat D_\mu\chi_-  -\bar\epsilon_-\gamma^{a0}e^\mu{}_a\hat D_\mu\chi_-
          +\frac2w\,\bar\epsilon_-\chi_-\,F_2  +\frac2w\,\bar\epsilon_-\gamma^0\chi_-\,F_1 \\
   &\quad -\frac14\,\bar\epsilon_+\gamma^{a0}\psi_{\mu-}\,e^\mu{}_a\,F_1
          -\frac14\,\bar\epsilon_+\gamma^a\psi_{\mu-}\,e^\mu{}_a\,F_2
          -\frac{w}4\,\bar\epsilon_+\gamma^0\phi_\mu\,e^\mu{}_a \\
   &\quad -\frac12\,\bar\epsilon_+\gamma^a\gamma^b\chi_-\,e^\mu{}_a\omega_\mu{}^b
          -\frac{w}6\,\bar\epsilon_+\gamma^{a0}\psi_{\mu+}\,e^\mu{}_a\tau^\rho\,r_\rho
          -\frac{w}6\,\bar\epsilon_+\gamma^a\gamma^b\psi_{\mu+}\,e^\mu{}_ae^\rho{}_b\,r_\rho \\
   &\quad -\frac{2(w+1)}{3}\,\bar\epsilon_+\gamma^0\chi_-\,\tau^\mu\,r_\mu
          -\frac{2(w+1)}{3}\,\bar\epsilon_-\gamma^a\chi_-\,e^\mu{}_a\,r_\mu
          -\frac{w}4\,\bar\epsilon_+\gamma^a\gamma^{b0}\psi_{\mu-}\,e^\mu{}_a\,b_b \\
   &\quad -(w+1)\,\bar\epsilon_-\gamma^{a0}\chi_-\,b_a
          +\frac14\,\bar\epsilon_+\gamma^a\gamma^b\chi_-\,\bar\psi_{\rho-}\gamma^b\psi_{\mu+}\,e^\mu{}_a\tau^\rho \\
   &\quad +\frac14\,\bar\epsilon_+\gamma^0\chi_-\,\bar\psi_{\rho-}\psi_{\mu-}\,e^\mu{}_ae^\rho{}_a
          +\frac14\,\bar\epsilon_+\gamma^{ab}\chi_-\,\bar\psi_{\rho-}\gamma^0\psi_{\mu-}\,e^\mu{}_ae^\rho{}_b \,.
\end{split}\end{align}
These are only the transformations of the independent fields. Those of the dependent gauge fields $\omega_\mu{}^{ab}$,
$\omega_\mu{}^a$,
$f_\mu$, $b_a$ and $\phi_\mu$ would be even longer, which is why we refrain from giving them here. They can be derived easily
from eqs.~\eqref{bostrafo}, \eqref{fermtrafo}, \eqref{depfieldbostrafos} and \eqref{depfieldfermtrafos}. Note that in the
transformations
of $\omega_\mu{}^a$ and $\phi_\mu$ one should also take into account the new expressions for curvatures of the gravitini
$\psi_{\mu-}$ and of $r_\mu$, see also the next section were we do work out those transformations for the dependent fields.

\subsection{``New minimal'' formulation \label{app:new}}

In the new minimal formulation the bosonic transformations of the dependent gauge fields $\omega_\mu{}^{ab}$, $\omega_\mu{}^a$,
$b_a$ and $\phi_\mu$ are given by
\begin{align}\begin{split}
 \delta \omega_\mu{}^{ab} &= \partial_\mu\lambda^{ab} \,, \\
 \delta \omega_\mu{}^a &= \partial_\mu\lambda^a -\omega_\mu{}^{ab}\lambda^b +\lambda^a\,e_\mu{}^b\,b_b +e_\mu{}^a \lambda^b\,b_b
                         +\lambda^a{}_b\,\omega_\mu{}^b \,, \\
 \delta b_a &= \lambda^a{}_b\,b_b \,, \\
 \delta \phi_\mu &= \frac14\,\lambda^{ab}\gamma_{ab}\phi_\mu -\gamma_0\phi_\mu\,\rho -\psi_{\mu+}\,\lambda^a\,b_a \,,
\end{split}\end{align}
while the fermionic transformations read
\begin{align}\begin{split}\label{depfieldtrafotor}
 \delta \omega_\mu{}^{ab} &= -\frac14\,\bar\epsilon_+\gamma^{ab0}\phi_\mu
               +\frac1{2w}\,\bar\epsilon_+\gamma^{ab}\psi_{\mu+}\,S
               +\bar\epsilon_-\gamma^{[a}\psi_{\mu+}\,b^{b]} \,, \\
 \delta \omega_\mu{}^a &= \bar\epsilon_-\gamma^0\hat\psi_\mu{}^a{}_- +\frac14\,e_\mu{}^b\,\bar\epsilon_+\gamma^b\hat\psi^a{}_{0-}
               +\frac14\,\bar\epsilon_+\gamma^a\hat\psi_{\mu0-}
               -\frac1w\,\bar\epsilon_+\gamma^{a0}\psi_{\mu-}\,S -\frac1w\,\bar\epsilon_-\gamma^{a0}\psi_{\mu+}\,S \\
  &\quad      -2\,\varepsilon^{ab}\,\bar\epsilon_-\psi_{\mu-}\,b_b
              +e_\mu{}^be^\rho{}_a\,\bar\epsilon_-\gamma^0\gamma^b\gamma^c\psi_{\rho-}\,b_c
              -\frac12\,e_\mu{}^be^\rho{}_a\,\bar\epsilon_-\gamma^b\big(\phi_\rho +\frac2w\,\gamma_0\psi_{\rho+}\,S\big) \\
  &\quad +\frac12\,e_\mu{}^a\,\tau^\rho\,\bar\epsilon_+\gamma^b\psi_{\rho+}\,b_b \,, \\
 \delta b_a &= -\frac12\,\bar\epsilon_+\gamma^b\psi_{\mu-}\,e^\mu{}_b\, b_a
               -\frac12\,\bar\epsilon_+\gamma^0\psi_{\mu+}\,\tau^\mu \, b_a
               -\frac14\,\bar\epsilon_+\gamma^0\phi_\mu\,e^\mu{}_a -\frac1{2w}\,\bar\epsilon_+\psi_{\mu+}\,e^\mu{}_a\,S \,, \\
 \delta \phi_\mu &= \epsilon_+\,f_\mu -\frac23\,\gamma^0\epsilon_+\,\big[\hat R_{\mu0}(R)
                        +\frac32\,\tau^\nu\,\bar\psi_{[\mu-}\gamma^{a0}\psi_{\nu]+}\,b_a
                        -\frac3{4w}\,\tau^\nu\,\bar\psi_{[\mu+}\gamma^0\psi_{\nu]+}\,S\big] \\
  &\quad +\frac43\,\gamma^a\epsilon_-\,\big[\hat R_{\mu a}(R) +\frac32\,e^\nu{}_a\,\bar\psi_{[\mu-}\gamma^{b0}\psi_{\nu]+}\,b_b
                        -\frac3{4w}\,e^\nu{}_a\,\bar\psi_{[\mu+}\gamma^0\psi_{\nu]+}\,S\big] \\
  &\quad -\big(D_\mu +e_\mu{}^a\,b_a +r_\mu\,\gamma_0\big)\big(\frac2w\,\gamma_0\epsilon_+\,S +2\,\gamma^{b0}\epsilon_-\,b_b\big) \,.
\end{split}\end{align}
Here, we used the covariant Newton--Cartan curvatures of the independent gauge fields $\psi_{\mu-}$ and $r_\mu$, which are are given
by
\begin{align}\begin{split}
 \hat\psi_{\mu\nu-} &= 2\,\partial_{[\mu}\psi_{\nu]-} -\frac12\,\omega_{[\mu}{}^{ab}\gamma_{ab}\psi_{\nu]-}
           -2\,r_{[\mu}\,\gamma_0\psi_{\nu]-} +\omega_{[\mu}{}^a\,\gamma_{a0}\psi_{\nu]+} \\
    &\quad +2\,\gamma^a\gamma^b\psi_{[\nu-}\,e_{\mu]}{}^ab_b +\frac2w\,\gamma_a\psi_{[\nu+}\,e_{\mu]}{}^a\,S \,, \\
 \hat R_{\mu\nu}(R) &= 2\,\partial_{[\mu}r_{\nu]} +\frac34\,\bar\psi_{[\mu+}\phi_{\nu]}
           -\frac32\,\bar\psi_{[\mu-}\gamma^{a0}\psi_{\nu]+}\,b_a +\frac3{4w}\,\bar\psi_{[\mu+}\gamma^0\psi_{\nu]+}\,S \,.
\end{split}\end{align}
Finally, the expression for the special conformal gauge field $f_\mu$ can be found  in eq.~\eqref{fmu}. We did not derive the
transformation rule of $f_\mu$ because no independent
field transforms to $f_\mu$. Therefore, its variation is not needed for any checks on the closure of the commutator algebra.


\small{


\providecommand{\href}[2]{#2}\begingroup\raggedright\endgroup

}

\end{document}